\documentclass[11pt,a4paper]{article} 

\newcommand{\PathSoft}{.}

\usepackage[a4paper,left=2cm, right=2cm, top=2.5cm, bottom=2.5cm]{geometry}

\usepackage[utf8]{inputenc} 
\usepackage{lmodern}
\usepackage[T1]{fontenc}

\usepackage[english]{babel} 
\usepackage{csquotes}

\usepackage{graphicx}
\usepackage{caption}
\usepackage{color}

\usepackage{amsmath, amsthm}
\usepackage{amssymb}

\usepackage{booktabs}

\usepackage{tikz}
\usetikzlibrary{arrows,shapes,positioning,calc}
\usepackage{pgfplots}

\definecolor{myred}{RGB}{112,25,25}
\definecolor{myblue}{RGB}{25,25,112}
\definecolor{mygreen}{RGB}{25,112,25}

\pgfplotsset{colormap={bluewhite}{ color(0cm)=(myblue!50!blue); color(1cm)=(white)}}
\pgfplotsset{colormap={redwhite}{ color(0cm)=(myred!50!red); color(1cm)=(white)}}
\pgfplotsset{colormap={greenwhite}{ color(0cm)=(mygreen!75!green); color(1cm)=(white)}}

\usepackage{natbib}
\usepackage[
    colorlinks,
    linkcolor=myblue!80!blue,
    citecolor=myblue!80!blue,
    urlcolor=myblue!60!blue
    ]{hyperref}

\usepackage{placeins} 
\usepackage{tabularx}
\usepackage{textcomp} 

\usepackage{framed}

\usepackage{Sweave}

\DefineVerbatimEnvironment{Sinput}{Verbatim}{fontsize=\small}
\DefineVerbatimEnvironment{Soutput}{Verbatim}{fontsize=\small}
\DefineVerbatimEnvironment{Scode}{Verbatim}{fontsize=\small}
\fvset{listparameters={\setlength{\topsep}{0pt}}}
  \renewenvironment{Schunk}{%
    \setlength{\topsep}{0pt}
    \colorlet{shadecolor}{black!7!white}
    \begin{snugshade*}}%
    {\end{snugshade*}}

\usepackage{bm} 
\usepackage{dsfont} 

\newcommand{\IR}{{\mathbb R}}
\newcommand{\IN}{{\mathbb N}}

\newcommand{\calT}{\mathcal{T}}

\newcommand{\N}{\operatorname{N}}

\newcommand{\iidsim}{\stackrel{\text{iid}}{\sim}}

\newcommand{\Var}{\operatorname{Var}} 
\newcommand{\IE}{{\mathbb E}}

\newcommand{\h}[1]{^{(#1)}}

\newcommand{\usw}{, \ldots ,}
\newcommand{\drm}{\mathrm{d}}

\newcommand{\inv}{^{-1}}

\newcommand{\trans}{^\top}

\newcommand{\mynorm}[1]{{\left\vert\kern-0.25ex\left\vert\kern-0.25ex\left\vert #1      \right\vert\kern-0.25ex\right\vert\kern-0.25ex\right\vert}}  

\newcommand{\vecnorm}[1]{{\left\vert\kern-0.25ex\left\vert #1      \right\vert\kern-0.25ex\right\vert}}
\newcommand{\myscal}[2]{\langle \kern-0.25ex \langle #1, #2\rangle \kern-0.25ex \rangle}
\newcommand{\norm}[1]{{\left\vert\kern-0.25ex\left\vert #1      \right\vert\kern-0.25ex\right\vert_2}}   
\newcommand{\scal}[2]{{\langle #1, #2\rangle_2}}

\newcommand{\pkg}[1]{{\bfseries #1}}
\newcommand{\code}[1]{{\ttfamily #1}}
\newcommand{\proglang}[1]{{\bfseries #1}}

\usepackage{authblk}

\begin{document}

\title{Object-Oriented Software for Functional Data}
\author{Clara Happ}
\date{}
\affil[1]{Department of Statistics, LMU Munich, Munich, Germany.}

\maketitle

\begin{abstract}
This paper introduces the \pkg{funData} \proglang{R} package as an object-oriented implementation of functional data. It implements a unified framework for dense univariate and multivariate functional data on one- and higher dimensional domains as well as for irregular functional data. The aim of this package is to provide a user-friendly, self-contained core toolbox for functional data, including important functionalities for creating, accessing and modifying functional data objects, that can serve as a basis for other packages. The package further contains a full simulation toolbox, which is a useful feature when implementing and testing new methodological developments.

Based on the theory of object-oriented data analysis, it is shown why it is natural to implement functional data in an object-oriented manner. The classes and methods provided by \pkg{funData} are illustrated in many examples using two freely available datasets.
The \pkg{MFPCA} package, which implements multivariate functional principal component analysis, is presented as an example for an advanced methodological package that uses the \pkg{funData} package as a basis, including a case study with real data.
Both packages are publicly available on GitHub and CRAN.
\end{abstract}

\section{Introduction}

Functional data analysis is  a branch of modern statistics that has seen a rapid growth in recent years.  The technical progress in many fields of application allows to collect data in increasingly fine resolution, e.g.,\ over time or space, such that the observed datapoints form quasi-continuous, possibly noisy, samples of smooth functions and are thus called \textit{functional data}. One central aspect of functional data analysis is that the focus of the analysis is not a single data point, but the entirety of all datapoints that are considered to stem from the same curve. Researchers in functional data analysis have developed many new statistical methods for the analysis of this type of data, linking the concept of functional data also to related branches of statistics, such as the study of longitudinal data, which can be seen as sparse and often also irregular samples of smooth functions, or image data, that can be represented as functions on two-dimensional domains. New approaches focus on even more generalized functional objects \citep[\textit{next generation} functional data analysis,][]{WangEtAl:2016}. 

When it comes to the practical application of new methods to real data, appropriate software solutions are needed to represent functional data in an adequate manner and ideally in a way that new theoretical developments can be implemented easily.
The most widely used \proglang{R} package for functional data is \pkg{fda} \citep{fda}, which is related to the popular textbook of \citet{RamsaySilverman:2005}. There are many other \proglang{R} packages for functional data that build on it, e.g.,\ \pkg{Funclustering} \citep{Funclustering}, \pkg{funFEM} \citep{funFEM} or \pkg{funHDDC}   \citep{funHDDC} or provide interfaces to \pkg{fda}, e.g.,\ \pkg{fda.usc} \citep{fda.usc} or \pkg{refund} \citep{refund}. The \pkg{fda} package contains a class \code{fd} for representing dense functional data on one-dimensional domains together with many functionalities for \code{fd} objects, such as plotting or summaries. It implements a variety of functional data methods, for example principal component analysis, regression models or registration. The \code{fd} objects represent the data as a finite linear combination of given basis functions, such as splines or Fourier bases, i.e., they store the basis functions and the individual coefficients for each curve. This representation of course works best if the underlying function is smooth and can be represented well in the chosen basis. Moreover, the data should be observed with only a small degree of noise.

Alternatively to the basis function representation, the raw, observed data can be saved directly. There are two different approaches for organizing the observations: Many packages use matrices, that contain the data in a row-wise (e.g.,\  \pkg{fda.usc}, \pkg{refund}) or column-wise (e.g.,\ \pkg{rainbow}, \citet{rainbow}) manner.   
This representation is most suitable for rather densely sampled data, where missing values can be coded via \code{NA}, which is supported by most of the packages. When it comes to irregular data, this way of storing functional data becomes quite inefficient, as the matrices then contain mostly missing values. Alternative solutions for sparse data or single points in 2D are list solutions (e.g.,\ \pkg{fdapace}, \citet{fdapace}) or \code{data.frame} based versions containing the data in a long format (e.g.,\ \pkg{fpca}, \citet{fpca}, \pkg{fdaPDE}, \citet{fdaPDE} or \pkg{sparseFLMM}, \citet{sparseFLMM}). Some packages also accept different formats (\pkg{funcy}, \citet{funcy} or \pkg{FDboost}, \citet{FDboost}). 

Technically, realizations of functional data on one-dimensional domains can also be interpreted as multivariate time series. The CRAN taks view for time series analysis (\url{https://CRAN.R-project.org/view=TimeSeries}) lists a lot of packages for this type of data, among which the \pkg{zoo} package \citep{zoo, ZeileisGrothendieck:2005} provides global infrastructure for regular and irregular time series. The main difference between functional data analysis and time series analysis is that for the former, each curve represents one observation of the same process, while for the latter the individual time points form the observations. Consequently, (multivariate) time series analysis aims more at analyzing the temporal dependence between curves and extrapolation/prediction of new time points, whereas the goal of functional data analysis is more in finding common structures between the curves (for example in PCA or clustering) and using them as predictors or response variables in regression models. More details on this topic can be found in the book of \citet{RamsaySilverman:2005}.

Image data, i.e., functions on two-dimensional domains are supported in \pkg{refund}, \pkg{refund.wave} \citep{refund.wave} and \pkg{fdasrvf}, \citep{fdasrvf}. Some others, as e.g.,\ \pkg{fda} and \pkg{fda.usc}  implement image objects, but use them rather for representing covariance or coefficient surfaces from function-on-function regression than for storing data in form of images. The majority of the \proglang{R} packages for functional data, however, are restricted to single functions on one-dimensional domains. 
Methods for multivariate functional data, consisting of more than one function per observation unit, have also become relevant in recent years. The \pkg{roahd} package \citep{roahd} provides a special class for this type of data, while some others simply combine the data from the different functions in a list (e.g.,\ \pkg{fda.usc}, \pkg{Funclustering} or \pkg{RFgroove}, \citet{RFgroove}). For all of these packages, the elements of the multivariate functional data must be observed on one-dimensional domains, which means that combinations of functions and images for example are not supported. In addition, the one-dimensional observation grid must be the same for most of the implementations.

In summary, there exist already several software solutions for functional data, but there  is still need for a unified, flexible representation of functional data, univariate and multivariate, on one- and higher dimensional domains and for dense and sparse functional data.
The \pkg{funData} package \citep{funData}, which is in the main focus of this article, attempts to fill this gap. It provides a unified framework to represent all these different types of functional data together with utility methods for handling the data objects. In order to take account of the particular structure of functional data, the implementation is organized in an object-oriented manner.
In this way, a link is established between the broad methodological field of object-oriented data analysis \citep{WangMarron:2007}, in which functional data analysis forms an important special case, and object-oriented programming \citep[e.g.,][]{Meyer:1988}, which is a fundamental concept in modern software engineering. It is shown why it is natural and reasonable to combine these two concepts for representing functional data.

In contrast to most \proglang{R} packages mentioned above, the \pkg{funData} package is not related to a certain type of methodology, such as regression, clustering or principal component analysis. Instead, it aims at providing a flexible and user-friendly core toolbox for functional data, which can serve as a basis for other packages, similarly to the \pkg{Matrix} package for linear algebra calculations for matrices \citep{Matrix}. It further contains a complete simulation toolbox for generating functional data objects, which is fundamental for testing new functional data methods.
The \pkg{MFPCA} package \citep{MFPCA}, which is also presented in this article,  is an example of a package that depends on \pkg{funData}. It implements a new methodological approach -- multivariate functional principal component analysis for data on potentially different dimensional domains \citep{HappGreven:2016} -- that allows to calculate principal components and individual score values for e.g.,\ functions and images, taking covariation between the elements into account. All implementation aspects that relate to functional data, i.e., input data, output data and all calculation steps involving functions are implemented using the object-oriented functionalities of the  \pkg{funData} package. Both packages are publicly available on GitHub (\url{https://github.com/ClaraHapp}) and CRAN (\url{https://CRAN.R-project.org}).

The structure of this article is as follows:
Section~\ref{sec:objectOrientation} contains a short introduction to the concept of object orientation in statistics and computer science and discusses how to adequately represent functional data in terms of software objects. The next section presents the object-oriented implementation of functional data in the \pkg{funData} package. Section~\ref{sec:MFPCApack} introduces the \pkg{MFPCA} package as an example on how to use the \pkg{funData} package for the implementation of new functional data methods. The final section contains a discussion and an outlook to potential future extensions.

\section{Object orientation and functional data}
\label{sec:objectOrientation}

Concepts of object orientation exist both in computer science and statistics. 
In statistics, the term \textit{object-oriented data analysis} (OODA) has been introduced by \citet{WangMarron:2007}. They define it as ``the statistical analysis of complex objects'' and draw their attention on what they call the \textit{atom} of the analysis. While in many parts of statistics these atoms are numbers or vectors (multivariate analysis), \citet{WangMarron:2007} argue that they can be much more complex objects such as images, shapes, graphs or trees. Functional data analysis \citep{RamsaySilverman:2005} is an important special case of object-oriented data analysis, where the atoms are functions. In most cases, they can be assumed to lie in $L^2(\calT)$, the space of square integrable functions on a domain $\calT$. This space has infinite dimension, but being a Hilbert space, its mathematical structure has many parallels to the space $\IR^p$ of $p$-dimensional vectors, which allows to transfer many concepts of multivariate statistics to the functional case in a quite straightforward manner.

In computer science, \textit{object orientation} \citep{Booch:2007, Armstrong:2006, Meyer:1988} is a programming paradigm which has profoundly changed the way how software is structured and developed. The key concept of object-oriented programming (OOP) is to replace the until then predominant procedural programs by computer programs made of objects, that can interact with each other and thus form, in a way, the ``atoms'' of the program. These objects usually consist of two blocks. First, a collection of data, which may have different data types, such as numbers, character strings or vectors of different length and is organized in fields. Second, a collection of methods, i.e., functions for accessing and/or modifying the data and for interacting with other objects. The entirety of all objects and their interactions forms the final program.

The main idea of the \pkg{funData} package is to combine the concepts of object orientation that exist in computer science and in statistics for the representation of functional data. The atom of the statistical analysis should thus be represented by the ``atom'' of the software program. The package therefore provides classes to organize the observed data in an appropriate manner. The class methods implement functionalities for accessing and modifying the data and for interaction between objects, which are primarily mathematical operations. 
The object orientation is realized in \proglang{R}  via the S4 object system \citep{Chambers:2008}. This system fulfills most of the fundamental concepts of object-oriented programming listed in \citet{Armstrong:2006} and is thus more rigorous than \proglang{R}'s widely used S3 system, which is used e.g.,\ by \pkg{fda} or \pkg{fda.usc}. In particular, it checks for example if a given set of observation (time) points matches the observed data before constructing the functional data object.

For the theoretical analysis of functional data, the functions are mostly considered as elements of a function space such as $L^2(\calT)$. For the practical analysis, the data can of course only be obtained in finite resolution. 
Data with functional features therefore will always come in pairs of the form $(t_{ij}, x_{ij})$ with \[x_{ij} = x_i(t_{ij}), \quad j = 1 \usw S_i,~i = 1 \usw N \]
for some functions $x_1\usw x_N$ that are considered as realizations of a random process $X: \calT \to \IR$. The domain $\calT \subset \IR^d$ here is assumed to be a compact set with finite (Lebesgue-) measure and in most cases, the dimension $d$ will be equal to $1$ (functions on one-dimensional domains), sometimes also $2$ (images) or $3$ (3D images). The observation points $t_{ij} \in \calT$ in general can differ in their number and location between the individual functions. 

When implementing functional data in an object-oriented way, it is thus natural to collect the data in two fields: the observation points $\{(t_{i1} \usw t_{iS_i}) \colon ~ i = 1 \usw N \}$ on one hand and the set of observed values   $\{(x_{i1} \usw x_{iS_i}) \colon ~ i = 1 \usw N \}$  on the other hand. Both fields form the data block of the functional data object as an inseparable entity.  This is a major advantage compared to non object-oriented implementations, that can consider the observation points  and the observed values as parameters in their methods, but can not map the intrinsic dependence between both of them.

In the important special case that the functions are observed on a one-dimensional domain and that the arguments do not differ across functions, they can be collected in a single vector $(t_1 \usw t_S)$ and the observed values can be stored in a matrix $X$ with entries $x_{ij},~i = 1\usw N,~ j = 1 \usw S$. 
The matrix-based concept can be generalized to data observed on common grids on higher dimensional domains. In this case, the observation grid can be represented as a matrix (or array) or, in the case of a regular and rectangular grid, as a collection of vectors that define the marginals of the observation grid. The observed data is collected in an array with three or even higher dimensions.

In recent years, the study of multivariate functional data that takes multiple functions at the same time into account, has led to new insights. Each observation unit here consists of a fixed number of functions  $p$, that can also differ in their domain \citep[e.g.,\ functions and images,][]{HappGreven:2016}. Technically, the observed values are assumed to stem from a random process $X = (X\h{1} \usw X\h{p})$, with random functions $X\h{k} \colon \calT_k \to \IR,~\calT_k \in \IR^{d_k},~ k = 1\usw p$, that we call the \textit{elements} of $X$. Realizations $x_1 \usw x_N$ of such a process all have the same structure as $X$. If for example $p = 2$ and $d_1 = 1,~d_2 = 2$, the realizations will all be bivariate functions with one functional and one image element. As data can only be obtained in finite resolution, observed  multivariate functional data is of the form
\[(t_{ij}\h{k},~ x_{ij}\h{k}) \quad j = 1 \usw S_i\h{k},~i = 1 \usw N,~ k = 1 \usw p.\]
Each element thus can be represented separately by its observation points and the observed  values, and the full multivariate sample constitutes the collection of all the $p$ elements.

\section[The funData package]{The \pkg{funData} package}
\label{sec:funData}

The  \pkg{funData} package implements the object-oriented approach for representing functional data in \proglang{R}. It provides three classes for functional data on one- and higher dimensional domains, multivariate functional data and irregularly sampled data, which are presented in Section~\ref{sec:Classes}. Section~\ref{sec:Methods} presents the methods associated with the functional data classes based on two example datasets and Section~\ref{sec:simTools} contains details on the integrated simulation toolbox.

\subsection{Three classes for functional data}
\label{sec:Classes}
For the representation of functional data in terms of abstract classes -- which, in turn, define concrete objects -- we distinguish three different cases.
\begin{enumerate}
\item \label{it:denseData} Class \code{funData} for dense functional data of ``arbitrary'' dimension (in most cases the dimension of the domain is $d \in \{1,2,3\}$) on a common set of observation points $t_1 \usw t_S$ for all curves. The curves may have missing values coded by \code{NA}.
\item \label{it:irregData} Class \code{irregFunData} for irregularly sampled  functional data  with individual sampling points $t_{ij},~ j = 1\usw S_i,~ i = 1 \usw N$ for all curves. The number $S_i$ and the location of observation points  can vary across individual observations. At the moment, only data on one-dimensional domains is implemented.
\item Class \code{multiFunData} for multivariate functional data, which combines $p$ elements of functional data that may be defined on different dimensional domains (e.g.,\ functions and images). 
\end{enumerate}

In the case of data on one-dimensional domains, the boundaries between the \code{funData} and the \code{irregFunData} class may of course be blurred in practice. The conceptual difference is that in~\ref{it:denseData}. all curves are ideally supposed to be sampled on the full grid $T = \{t_1 \usw t_S\} \subset \calT$ and differences in the number of observation points per curve are mainly driven by anomalies or errors in the sampling process, such as missing values, which can be coded by \code{NA}. In contrast, case~\ref{it:irregData}. \textit{a priori} expects that the curves can be observed at different observation points $t_{ij}$, and that the number of observations per curve may vary.

\begin{figure}
\centering
\begin{minipage}[t][][c]{0.35\textwidth}
\includegraphics[width = \textwidth, page = 1]{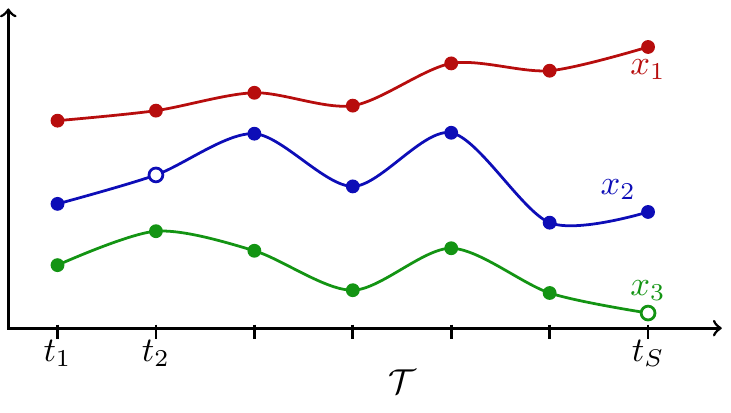}
\end{minipage}
\hspace*{0.5cm}
\begin{minipage}[t][][c]{0.4\textwidth}
\includegraphics[width = \textwidth, page = 2]{figures/plotAll.pdf}
\end{minipage}
\caption{Left: An example of $N=3$ observations of functional data on a one-dimensional domain $\calT$, observed on a common discrete grid $(t_1 \usw t_S)$, where the observed values $x_{ij} = x_i(t_j)$ are represented by solid circles. The functions $x_2$ and $x_3$ have one missing value, each (open circles). Right: Representation of the data in a \code{funData} object. The \code{@argvals} slot is a list of length one, containing the observation grid as a vector. The \code{@X} slot is a matrix of dimension $N \times S$ that contains the observed values in row-wise format. Missing values are coded with \code{NA}.}
\label{fig:funData_1D}
\end{figure}

\begin{figure} 
\centering 
\begin{minipage}[b][][c]{0.57\textwidth}
\includegraphics[width = 0.32\textwidth, page = 3]{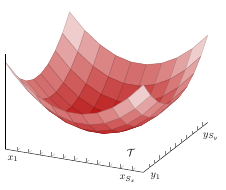}
\includegraphics[width = 0.32\textwidth, page = 4]{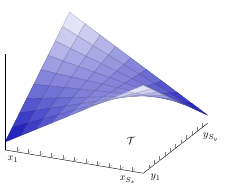}
\includegraphics[width = 0.32\textwidth, page = 5]{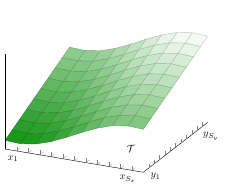}
\end{minipage}
\hspace*{0.1cm}
\begin{minipage}[b][][c]{0.4\textwidth}
\includegraphics[width = \textwidth, page = 6]{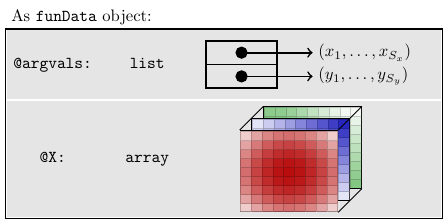}
\end{minipage}
\caption{Left: An example of $N = 3$ observations of functional data on a two-dimensional domain $\calT$. The functions are observed on a common discrete grid having $S_x$ points in $x$- and $S_y$ points in $y$-direction, i.e., each observation forms an image with $S_x \times S_y$ pixels.
Right: Representation of the data in a \code{funData} object. The \code{@argvals} slot is a list of length $2$, containing the marginal sampling points. The slot \code{@X} is an array of dimension $N \times S_x \times S_y$.}
\label{fig:funData_2D}
\end{figure}

\begin{figure}   \centering
\begin{minipage}[t][][c]{0.35\textwidth}
\includegraphics[width = \textwidth, page = 7]{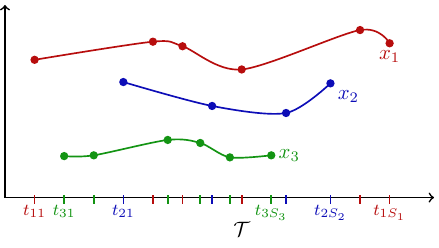}
\end{minipage}
\hspace*{0.5cm}
\begin{minipage}[t][][c]{0.4\textwidth}
\includegraphics[width = \textwidth, page = 8]{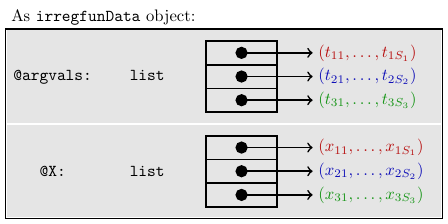}
\end{minipage}
\caption{Left: An example of $N=3$ irregular observations of functional data on a one-dimensional domain $\calT$. The observation points for each function differ in number and location.  Right: Representation of the data in an \code{irregFunData} object. Both the \code{@argvals} and the \code{@X} slot are a list of length $N$, containing the observation points $t_{ij}$ and the observed values $x_{ij}$.}
\label{fig:irregFunData}
\end{figure}

\begin{figure} 
\centering

\begin{minipage}[c][][c]{0.35\textwidth}
\includegraphics[width = \textwidth, page = 9]{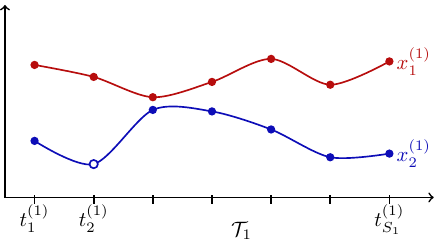}

\includegraphics[width = 0.45\textwidth, page = 11]{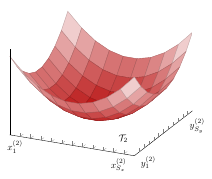}
\hfill
\includegraphics[width = 0.45\textwidth, page = 12]{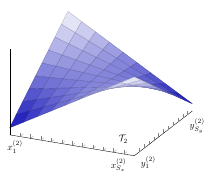}
\end{minipage}
\hspace*{0.5cm}
\begin{minipage}[c][][c]{0.5\textwidth}
\includegraphics[width = \textwidth, page = 14]{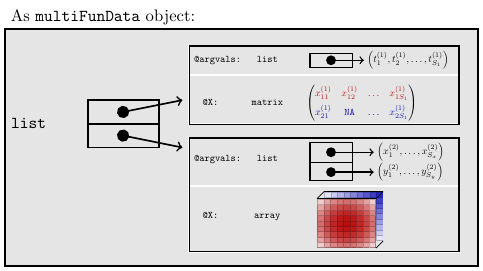}
\end{minipage}

\caption{Left: An example of $N=2$ observations of bivariate functional data on different domains, i.e., each observation (red/blue) consists of two elements, a curve and an image. Right: Representation of the data as a \code{multiFunData} object. As the data is bivariate, the \code{multiFunData} object is a list of length $2$, containing the two elements as \code{funData} objects.}
\label{fig:multiFunData}
\end{figure}

For \code{funData} and \code{irregFunData}, the data is organized in two fields or \textit{slots}, as they are called for S4 classes \citep{Chambers:2008}: The slot \code{@argvals} contains the observation points and the slot \code{@X} contains the observed data. For \code{funData}, the \code{@argvals} slot is a list, containing the common sampling grid for all functions and \code{@X} is an array containing all observations.
In the simplest case of functions defined on a one-dimensional domain and sampled on a grid with $S$ observation points, \code{@argvals} is a list of length one, containing a vector of length $S$ and \code{@X} is a matrix of dimension $N \times S$, containing the observed values for each curve in a row-wise manner. For an illustration, see Figure~\ref{fig:funData_1D}. 
If the \code{funData} object is supposed to represent $N$ images with $S_x \times S_y$ pixels, \code{@argvals} is a list of length 2, containing two vectors with $S_x$ and $S_y$ entries, respectively, that represent the sampling grid. The slot \code{@X} is an array of dimension $N \times S_x \times S_y$, cf. Figure~\ref{fig:funData_2D}. 
For the \code{irregFunData} class, only functions on one-dimensional domains are currently implemented. The \code{@argvals} slot here is a list of length $N$, containing in its $i$-th entry the vector $(t_{i1} \usw t_{iS_i})$ with all observation points for the $i$-th curve. The \code{@X} slot organizes the observed values analogously, i.e., it is also a list of length $N$ with the $i$-th entry containing the vector $(x_{i1} \usw x_{iS_i})$. An illustration is given in Figure~\ref{fig:irregFunData}. 
The \code{multiFunData} class, finally, represents multivariate functional data with $p$ elements. An object of this class is simply a list of $p$ \code{funData} objects, representing the different elements. For an illustration, see Figure~\ref{fig:multiFunData}. 
Given specific data, the realizations of such classes are called \code{funData}, \code{irregFunData} or \code{multiFunData} objects. We will use the term \textit{functional data object} in the following for referring to objects of all three classes.

\subsection{Methods for functional data objects}
\label{sec:Methods}

Essential methods for functional data  objects include for example the creation of an object from the observed data, methods for modifying and subsetting the data, plotting, arithmetics. The methods in the \pkg{funData} package are implemented such that they work on functional data objects as the atoms of the program, i.e., the methods accept functional data objects as input and/or have functional data objects as output. Moreover, all functions are implemented for the three different classes with appropriate sub-methods. This corresponds to the principle of polymorphism in \citet{Armstrong:2006}, as different classes have their own implementation e.g.,\ of a plot function. In most cases, the methods for \code{multiFunData} objects will simply call the corresponding method for each element and concatenate the results appropriately.

\subsubsection{Data used in the examples}
The following code examples use the Canadian weather data \citep{RamsaySilverman:2005}, that is available e.g.,\ in the \pkg{fda} package and the CD4 cell count data \citep{GoldsmithEtAl:2013} from the \pkg{refund} package. In both cases, the data is observed on a one-dimensional domain. Examples for image data are included in the description of the simulation toolbox (Section~\ref{sec:simTools}).

The Canadian weather data contains daily and monthly observations of temperature (in °C) and precipitation (in mm) for $N = 35$ Canadian weather stations, averaged over the years 1960 to 1994. We will use the daily temperature as an example for dense functional data on a one-dimensional domain. Moreover, it is combined with the monthly precipitation data to  multivariate functional data with elements on different domains ($\calT_1 = [1,365]$ for the temperature and $\calT_2 = [1,12]$ for the precipitation).

The CD4 cell count data reports the number of CD4 cells per milliliter of blood for $N = 366$ subjects who participated in a study on AIDS (MACS, Multicenter AIDS Cohort Study). CD4 cells are part of the human immune system and are attacked in the case of an infection with HIV. Their number thus can be interpreted as a proxy for the disease progression. For the present data, the CD4 counts were measured roughly twice a year and centered at the time of seroconversion, i.e., the time point when HIV becomes detectable. In total, the number of observations for each subject varies between $1$ and $11$ in the period of 18 months before and 42 months after seroconversion. The individual time points do also differ between subjects. The dataset thus serves as an example for irregular functional data. For more information on the data, please see \citet{GoldsmithEtAl:2013}.

\subsubsection{Creating new objects and accessing an object's information}
The following code creates \code{funData} objects for the Canadian temperature and precipitation data:
\begin{Schunk}
\begin{Sinput}
R> data("CanadianWeather", package = "fda")
R> dailyTemp <- funData(argvals = 1:365,
+                       X = t(CanadianWeather$dailyAv[,, 'Temperature.C']))
R> monthlyPrec <- funData(argvals = 1:12,
+                         X = t(CanadianWeather$monthlyPrecip))
\end{Sinput}
\end{Schunk}
It is then very easy to create a bivariate \code{multiFunData} object, containing the daily temperature and the monthly precipitation for the $35$ weather stations:
\begin{Schunk}
\begin{Sinput}
R> canadWeather <- multiFunData(dailyTemp, monthlyPrec)
\end{Sinput}
\end{Schunk}

The \code{cd4} data in the \pkg{refund} package is stored in a \code{matrix} with $366 \times 61$ entries, containing the CD4 counts for each patient on the common grid of all sampling points. Missing values are coded as \code{NA}. 
Since each patient has at least $1$ and at most $11$ observations, more than $90\%$ of the dataset consists of missings. Particularly, the time of seroconversion (time 0) is missing for all subjects. The \code{irregFunData} class stores only the observed values and their time points and is therefore  more parsimonious. The following code extracts both separately as lists and then constructs an \code{irregFunData} object:
\begin{Schunk}
\begin{Sinput}
R> data("cd4", package = "refund")
R> allArgvals <- seq(-18, 42)
R> argvalsList <- apply(cd4, 1, function(x){allArgvals[complete.cases(x)]})
R> obsList <- apply(cd4, 1, function(x){x[complete.cases(x)]})
R> cd4Counts <- irregFunData(argvals = argvalsList, X = obsList)
\end{Sinput}
\end{Schunk}

When a functional data object is called in the command line, some basic information is printed to standard output. For the \code{funData} object containing the Canadian temperature data:
\begin{Schunk}
\begin{Sinput}
R> dailyTemp
\end{Sinput}
\begin{Soutput}
Functional data with 35 observations of 1-dimensional support
argvals:
	1 2 ... 365		(365 sampling points)
X:
	array of size 35 x 365 
\end{Soutput}
\end{Schunk}
The \code{multiFunData} version lists the information of the different elements:
\begin{Schunk}
\begin{Sinput}
R> canadWeather
\end{Sinput}
\begin{Soutput}
An object of class "multiFunData"
[[1]]
Functional data with 35 observations of 1-dimensional support
argvals:
	1 2 ... 365		(365 sampling points)
X:
	array of size 35 x 365 

[[2]]
Functional data with 35 observations of 1-dimensional support
argvals:
	1 2 ... 12		(12 sampling points)
X:
	array of size 35 x 12 

\end{Soutput}
\end{Schunk}
For \code{irregFunData} objects there is some additional information about the total number of observations. Note that time 0 has been dropped here, as there are no observations.
\begin{Schunk}
\begin{Sinput}
R> cd4Counts
\end{Sinput}
\begin{Soutput}
Irregular functional data with 366 observations of 1-dimensional support
argvals:
	Values in -18 ... 42.
X:
	Values in 10 ... 3184.
Total:
	1888 observations on 60 different argvals (1 - 11 per observation).
\end{Soutput}
\end{Schunk}
More information can be obtained using the usual \code{summary} or \code{str} functions:
\begin{Schunk}
\begin{Sinput}
R> summary(dailyTemp)
\end{Sinput}
\end{Schunk}
\begin{Schunk}
\begin{Soutput}
Argument values (@argvals):
         Min. 1st Qu. Median Mean 3rd Qu. Max.
Dim. 1 :    1      92    183  183     274  365

Observed functions (@X):
        St. Johns Halifax Sydney Yarmouth Charlottvl Fredericton
Min.        -7.00   -8.10  -8.40   -5.300    -10.400     -12.400
1st Qu.     -2.10   -2.60  -2.40   -0.100     -3.800      -4.700
Median       4.50    6.40   5.40    7.400      5.700       6.300
Mean         4.69    6.15   5.51    6.812      5.232       5.263
  ... [output truncated] ... 

\end{Soutput}
\end{Schunk}
\begin{Schunk}
\begin{Sinput}
R> str(cd4Counts)
\end{Sinput}
\end{Schunk}
\begin{Schunk}
\begin{Soutput}
IrregFunData:
366 observations of 1-dimensional support on 60 different argvals (1 - 11 per curve).

@argvals: List of 366
 $ : int [1:3] -9 -3 3
 $ : int [1:4] -3 3 9 15
 $ : int [1:8] -15 -9 -3 3 9 17 22 29
  [list output truncated]

@X: List of 366
 $ : Named int [1:3] 548 893 657
  ..- attr(*, "names")= chr [1:3] "-9" "-3" "3"
 $ : Named int [1:4] 752 459 181 434
  ..- attr(*, "names")= chr [1:4] "-3" "3" "9" "15"
 $ : Named int [1:8] 846 1102 801 824 866 704 757 726
  ..- attr(*, "names")= chr [1:8] "-15" "-9" "-3" "3" ...
  [list output truncated]
\end{Soutput}
\end{Schunk}
The slots can be accessed directly via \code{@argvals} or \code{@X}. The preferable way of accessing and modifying the data in the slots, however, is via the usual get/set methods, following the principle of limited access \citep[or encapsulation,][]{Armstrong:2006}, as an example:
\begin{Schunk}
\begin{Sinput}
R> argvals(monthlyPrec)
\end{Sinput}
\begin{Soutput}
[[1]]
 [1]  1  2  3  4  5  6  7  8  9 10 11 12

\end{Soutput}
\end{Schunk}
The names can be set or get by the \code{names} function:
\begin{Schunk}
\begin{Sinput}
R> names(monthlyPrec) <- names(dailyTemp)
R> names(monthlyPrec)
\end{Sinput}
\begin{Soutput}
 [1] "St. Johns"   "Halifax"     "Sydney"      "Yarmouth"    "Charlottvl" 
 [6] "Fredericton" "Scheffervll" "Arvida"      "Bagottville" "Quebec"     
[11] "Sherbrooke"  "Montreal"   
 ... [output truncated] ... 
\end{Soutput}
\end{Schunk}
The method \code{nObs} returns the number of observations (functions) in each object:
\begin{Schunk}
\begin{Sinput}
R> nObs(dailyTemp)
\end{Sinput}
\begin{Soutput}
[1] 35
\end{Soutput}
\begin{Sinput}
R> nObs(cd4Counts)
\end{Sinput}
\begin{Soutput}
[1] 366
\end{Soutput}
\begin{Sinput}
R> nObs(canadWeather) 
\end{Sinput}
\begin{Soutput}
[1] 35
\end{Soutput}
\end{Schunk}
The number of observation points is given by \code{nObsPoints}. Note that the functions in \code{dailyTemp} and \code{canadWeather} are densely sampled and therefore return a single number or two numbers (one for each element). For the irregularly sampled data in \code{cd4Counts}, the method returns a vector of length $N = 366$, containing the individual number of observations for each subject:
\begin{Schunk}
\begin{Sinput}
R> nObsPoints(dailyTemp)
\end{Sinput}
\begin{Soutput}
[1] 365
\end{Soutput}
\begin{Sinput}
R> nObsPoints(cd4Counts)
\end{Sinput}
\begin{Soutput}
 [1] 3 4 8 4 8 3 4 7 2 6 8 3
 ... [output truncated] ... 
\end{Soutput}
\begin{Sinput}
R> nObsPoints(canadWeather) # 365 (temp.) and 12 (prec.) observation points
\end{Sinput}
\begin{Soutput}
[[1]]
[1] 365

[[2]]
[1] 12

\end{Soutput}
\end{Schunk}
The dimension of the domain can be obtained by the \code{dimSupp} method:
\begin{Schunk}
\begin{Sinput}
R> dimSupp(dailyTemp)
\end{Sinput}
\begin{Soutput}
[1] 1
\end{Soutput}
\begin{Sinput}
R> dimSupp(cd4Counts)
\end{Sinput}
\begin{Soutput}
[1] 1
\end{Soutput}
\begin{Sinput}
R> dimSupp(canadWeather) # both elements have one-dimensional support
\end{Sinput}
\begin{Soutput}
[1] 1 1
\end{Soutput}
\end{Schunk}
Finally, a subset of the data can be extracted using \proglang{R}'s usual bracket notation or via  the function \code{extractObs} (alias \code{subset}). We can for example extract the temperature data for the first five weather stations:
\begin{Schunk}
\begin{Sinput}
R> dailyTemp[1:5]
\end{Sinput}
\begin{Soutput}
Functional data with 5 observations of 1-dimensional support
argvals:
	1 2 ... 365		(365 sampling points)
X:
	array of size 5 x 365 
\end{Soutput}
\end{Schunk}
or the CD4 counts of the first 8 patients before seroconversion (i.e., until time $0$):
\begin{Schunk}
\begin{Sinput}
R> extractObs(cd4Counts, obs = 1:8, argvals = -18:0)
\end{Sinput}
\begin{Soutput}
Irregular functional data with 8 observations of 1-dimensional support
argvals:
	Values in -17 ... -3.
X:
	Values in 429 ... 1454.
Total:
	15 observations on 6 different argvals (1 - 3 per observation).
\end{Soutput}
\end{Schunk}
In both cases, the method returns an object of the same class as the argument with which the function was called (\code{funData} for \code{dailyTemp} and \code{irregFunData} for \code{cd4Counts}), which is seen by the output.

\subsubsection{Plotting}
\begin{figure}[ht]
\centering
\includegraphics[width = 0.35\textwidth]{\PathSoft/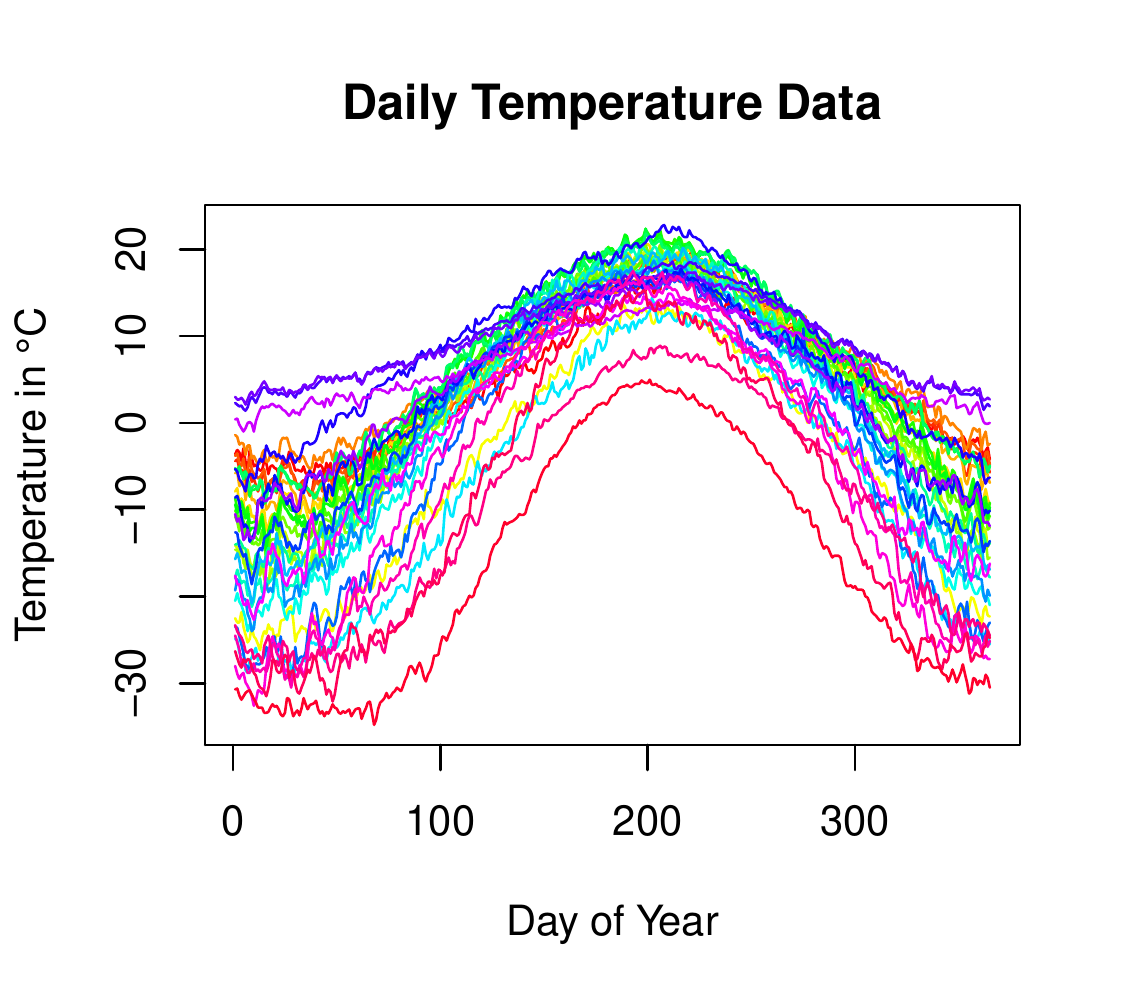}
\includegraphics[width = 0.35\textwidth]{\PathSoft/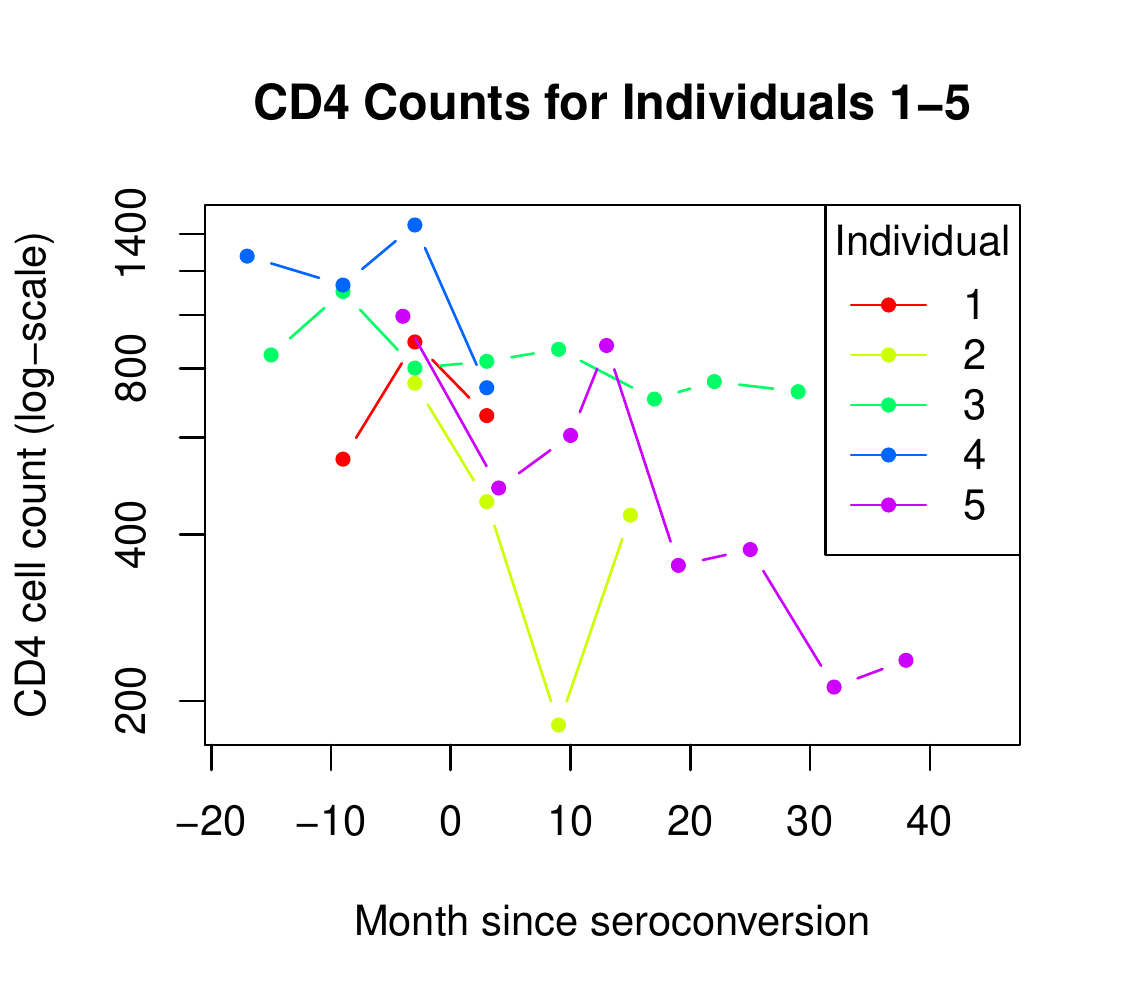}

\includegraphics[width = 0.7\textwidth]{\PathSoft/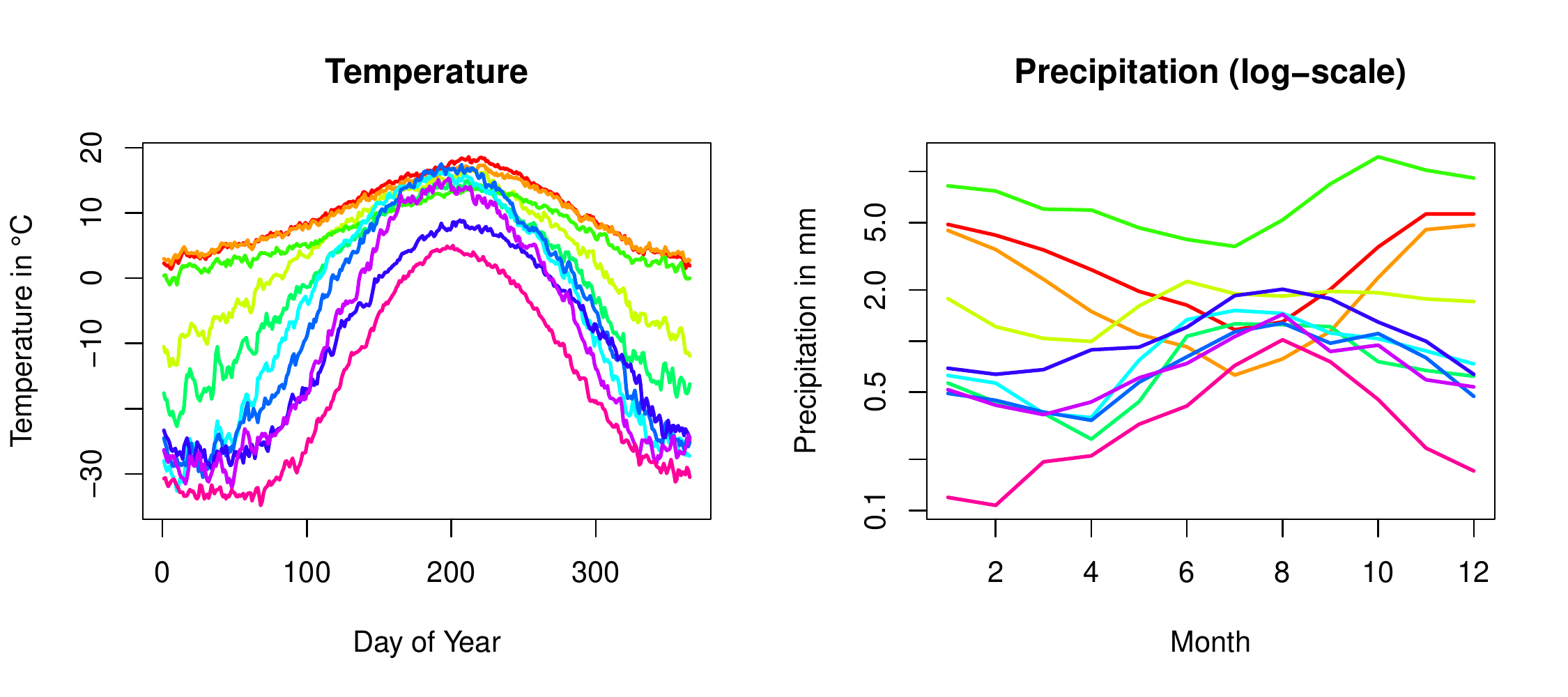}
\caption{Results of the \code{plot} commands for functional data objects. First row: The daily temperature in $35$ Canadian weather stations (\code{funData} object, left) and the CD4 counts for the first five individuals (\code{irregFunData} object, right). Second row: The Canadian weather data for ten weather stations (\code{multiFunData} object). See text for the commands used; all other options were kept as defaults.}
\label{fig:plot}
\end{figure}

The more complex the data, the more important it is to have adequate visualization methods. The \pkg{funData} package comes with two plot methods for each class, one based on \proglang{R}'s standard plotting engine (\code{plot.default}  and \code{matplot}) and one based on the \code{ggplot2} implementation of the grammar of graphics (package \pkg{ggplot2}, \citet{Wickham:2009, ggplot2}).
The \code{plot} function inherits all parameters from the \code{plot.default} function from the \pkg{graphics} package, i.e., colors, axis labels and many other options can be set as usual. The following code plots all 35 curves of the Canadian temperature data: 
\begin{Schunk}
\begin{Sinput}
R> plot(dailyTemp, main = "Daily Temperature Data",
+       xlab = "Day of Year", ylab = "Temperature in °C")
\end{Sinput}
\end{Schunk}
and the CD4 counts of the first five patients on the log-scale:
\begin{Schunk}
\begin{Sinput}
R> plot(cd4Counts, obs = 1:5, xlim = c(-18, 45), log = "y", 
+       main = "CD4 Counts for Individuals 1-5",
+       xlab = "Month since seroconversion", 
+       ylab = "CD4 cell count (log-scale)")
R> legend("topright", legend = 1:5, col = rainbow(5), lty = 1, pch = 20, 
+         title = "Individual")
\end{Sinput}
\end{Schunk}
For multivariate functional data, the different elements are plotted side by side, as shown here for the last ten Canadian weather stations:
\begin{Schunk}
\begin{Sinput}
R> plot(canadWeather, obs = 26:35, lwd = 2, log = c("", "y"), 
+       main = c("Temperature", "Precipitation (log-scale)"), 
+       xlab = c("Day of Year", "Month"),
+       ylab = c("Temperature in °C", "Precipitation in mm"))
\end{Sinput}
\end{Schunk}
The resulting plots are shown in Figure~\ref{fig:plot}.

\begin{figure}[ht]
   \centering
\hspace{-1cm}
\includegraphics[width = 0.25\textwidth]{\PathSoft/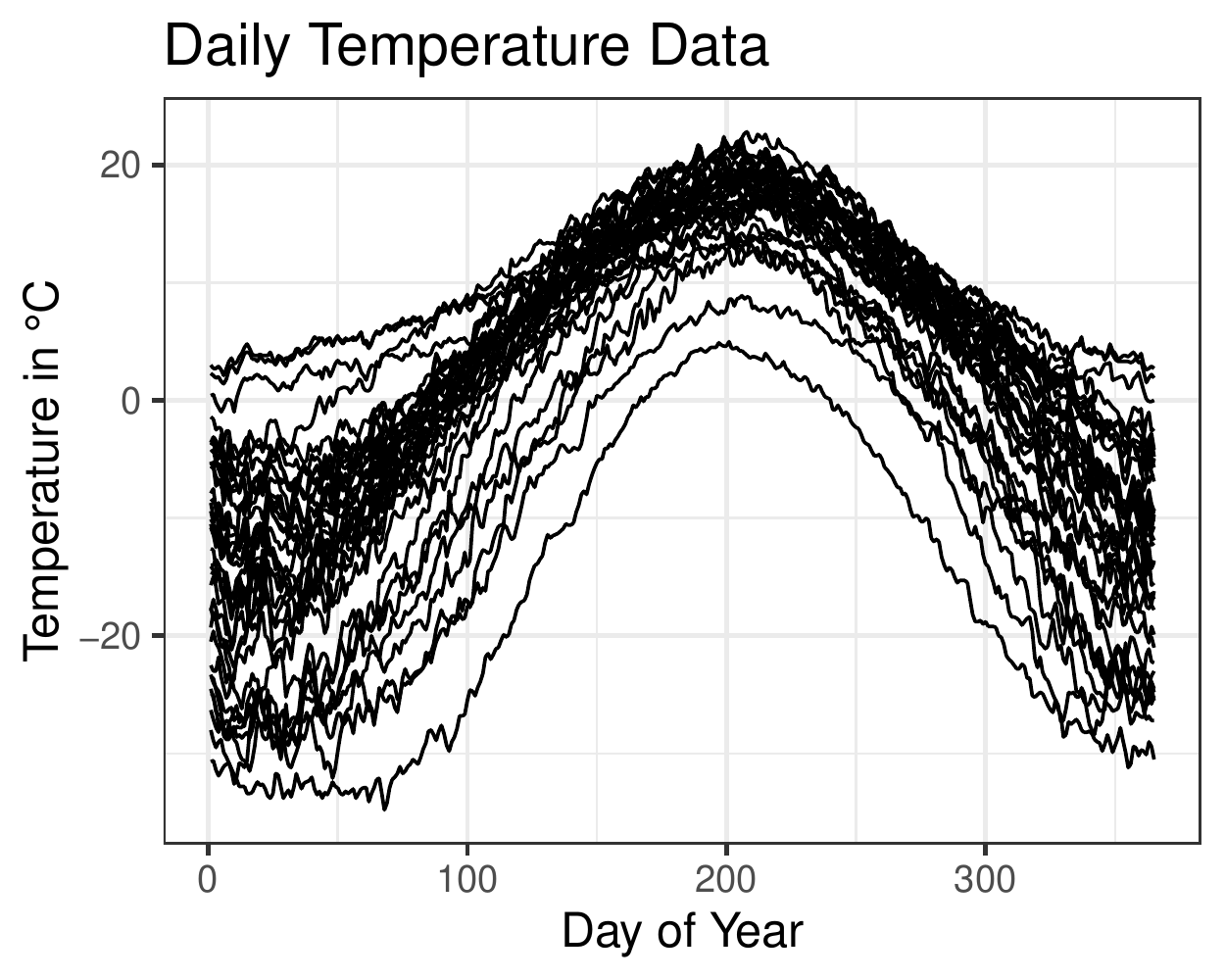}
\hspace{1.75cm}
\includegraphics[width = 0.3\textwidth]{\PathSoft/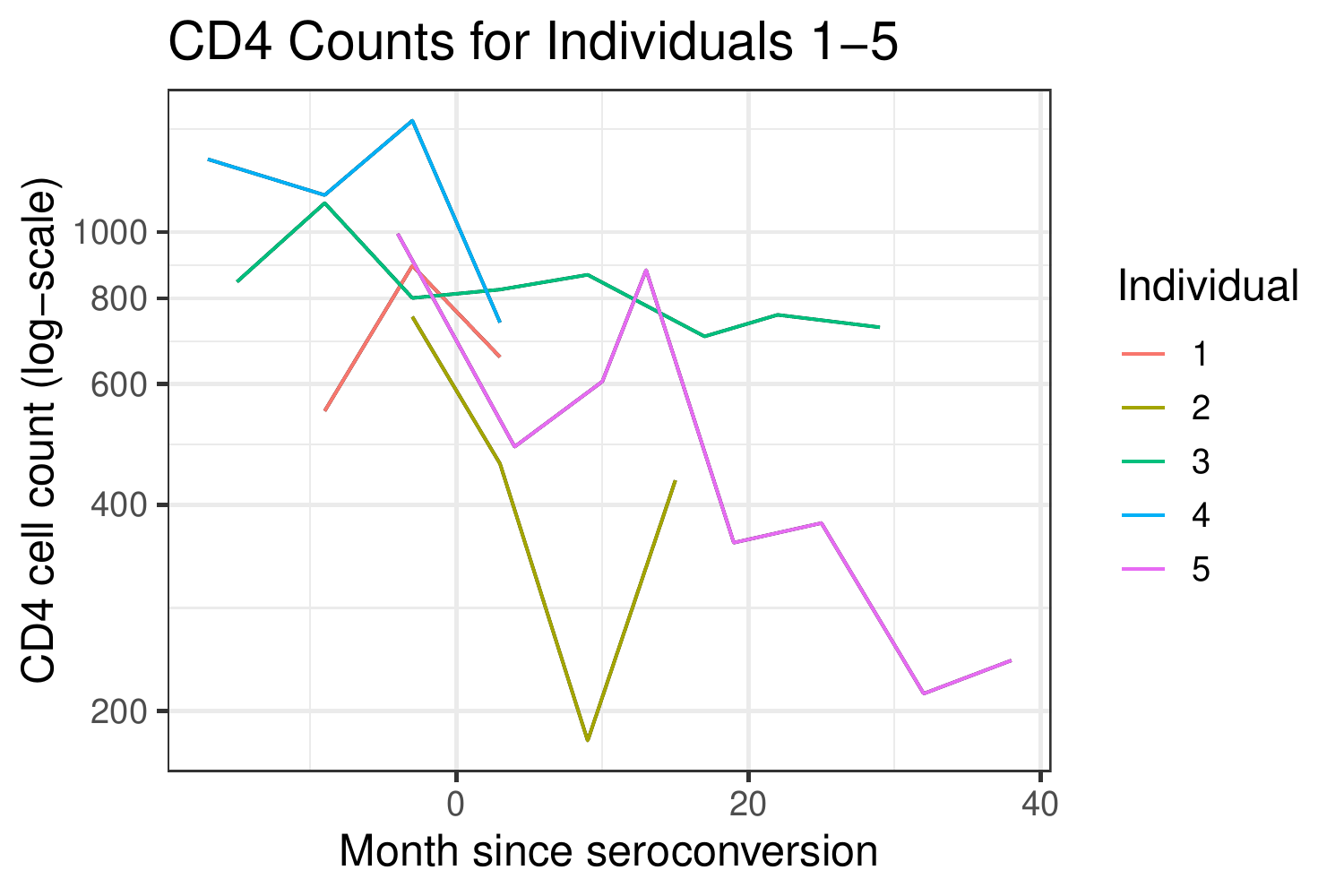}

\centering
\includegraphics[width = 0.7\textwidth]{\PathSoft/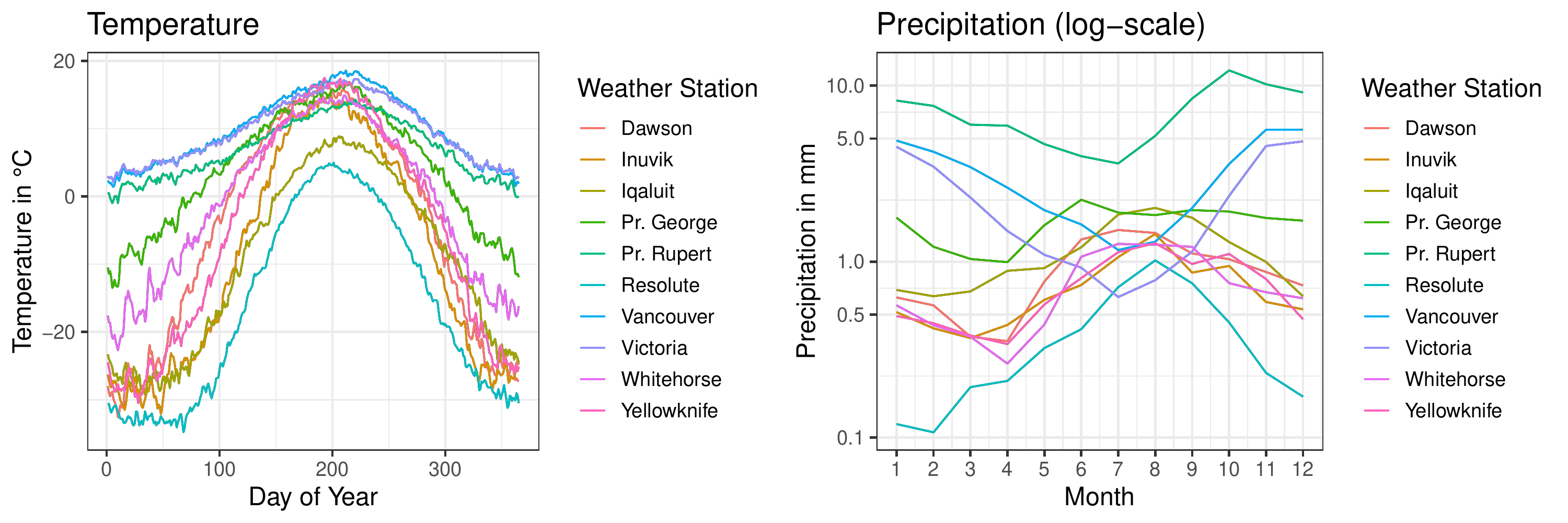}
\caption{Results of the \code{autoplot} commands for functional data objects. First row: The daily temperature in $35$ Canadian weather stations (\code{funData} object, left) and the CD4 counts for the first five individuals (\code{irregFunData} object, right). Second row: The Canadian weather data for ten weather stations (\code{multiFunData} object). See text for the commands used. For all plots the option \code{theme\_bw()} has been added for optimal print results; all other parameters were kept as defaults.}
\label{fig:ggplot}
\end{figure}

The optional \code{autoplot} / \code{autolayer} functions create a \code{ggplot} object that can be further modified by the user by loading the \pkg{ggplot2} package and using the functionality provided therein. The following codes produce analogous plots to the \code{plot} examples above for the Canadian temperature data:
\begin{Schunk}
\begin{Sinput}
R> library("ggplot2")
R> 
R> tempPlot <- autoplot(dailyTemp)
R> tempPlot + labs(title = "Daily Temperature Data", 
+                  x = "Day of Year", y = "Temperature in °C")
\end{Sinput}
\end{Schunk}
and for the CD4 counts:
\begin{Schunk}
\begin{Sinput}
R> cd4Plot <- autoplot(cd4Counts, obs = 1:5)
R> cd4Plot + geom_line(aes(colour = obs)) +
+    labs(title = "CD4 Counts for Individuals 1-5", color = "Individual",
+       x = "Month since seroconversion", y = "CD4 cell count (log-scale)") +
+    scale_y_log10(breaks = seq(200,1000,200))
\end{Sinput}
\end{Schunk}
For the bivariate Canadian weather data, the bivariate plot is obtained via:
\begin{Schunk}
\begin{Sinput}
R> weatherPlot <- autoplot(canadWeather, obs = 26:35)
\end{Sinput}
\end{Schunk}
The subplots of the different elements can be customized separately, by changing for example the colors of the curves or adding axis labels, using functions from the \pkg{ggplot2} package.
\begin{Schunk}
\begin{Sinput}
R> weatherPlot[[1]] <- weatherPlot[[1]] + geom_line(aes(colour = obs)) +
+    labs(title = "Temperature", colour = "Weather Station", 
+         x = "Day of Year", y = "Temperature in °C")
R> 
R> weatherPlot[[2]] <- weatherPlot[[2]] + geom_line(aes(colour = obs)) + 
+    labs(title = "Precipitation (log-scale)", colour = "Weather Station",
+         x = "Month", y = "Precipitation in mm") + 
+    scale_x_continuous(breaks = 1:12) + 
+    scale_y_log10(breaks = c(0.1,0.5,1,5,10))
\end{Sinput}
\end{Schunk}
For the final plot, the subplots are arranged side by side using the \pkg{gridExtra} package \citep{gridExtra}:
\begin{Schunk}
\begin{Sinput}
R> gridExtra::grid.arrange(grobs = weatherPlot, nrow = 1) 
\end{Sinput}
\end{Schunk}
The corresponding plots for all three data examples are shown in Figure~\ref{fig:ggplot}.

\FloatBarrier

\subsubsection{Coercion}
As discussed earlier, there is no clear boundary between the \code{irregFunData} class and the \code{funData} class for functions on one-dimensional domains. The package thus provides coercion methods to convert \code{funData} objects to \code{irregFunData} objects, as seen in the output:
\begin{Schunk}
\begin{Sinput}
R> as.irregFunData(dailyTemp)
\end{Sinput}
\begin{Soutput}
Irregular functional data with 35 observations of 1-dimensional support
argvals:
	Values in 1 ... 365.
X:
	Values in -34.8 ... 22.8.
Total:
	12775 observations on 365 different argvals (365 - 365 per observation).
\end{Soutput}
\end{Schunk}
Vice versa the union of all observation points of all subjects is used as the common one and missing values are coded with \code{NA} (\code{as.funData(cd4Counts)}).
Similarly, \code{funData} objects can also be coerced to \code{multiFunData} objects with only one element.

In order to simplify working with other \proglang{R} packages, functional data objects can be converted to a long data format via the function \code{as.data.frame}, here  exemplarily shown for the CD4 count data:
\begin{Schunk}
\begin{Sinput}
R> as.data.frame(cd4Counts)
\end{Sinput}
\begin{Soutput}
     obs argvals    X
1      1      -9  548
2      1      -3  893
3      1       3  657
4      2      -3  752
  ... [output truncated] ... 
\end{Soutput}
\end{Schunk}
The \pkg{funData} package further provides coercion methods between \code{funData} objects and \code{fd} objects from package \pkg{fda} (\code{funData2fd} and \code{fd2funData}), which provides analysis tools for functional data and is also the basis of many other \proglang{R} packages.

\subsubsection{Mathematical operations for functional data objects}
\begin{figure}   \centering
\includegraphics[width = 0.7\textwidth]{\PathSoft/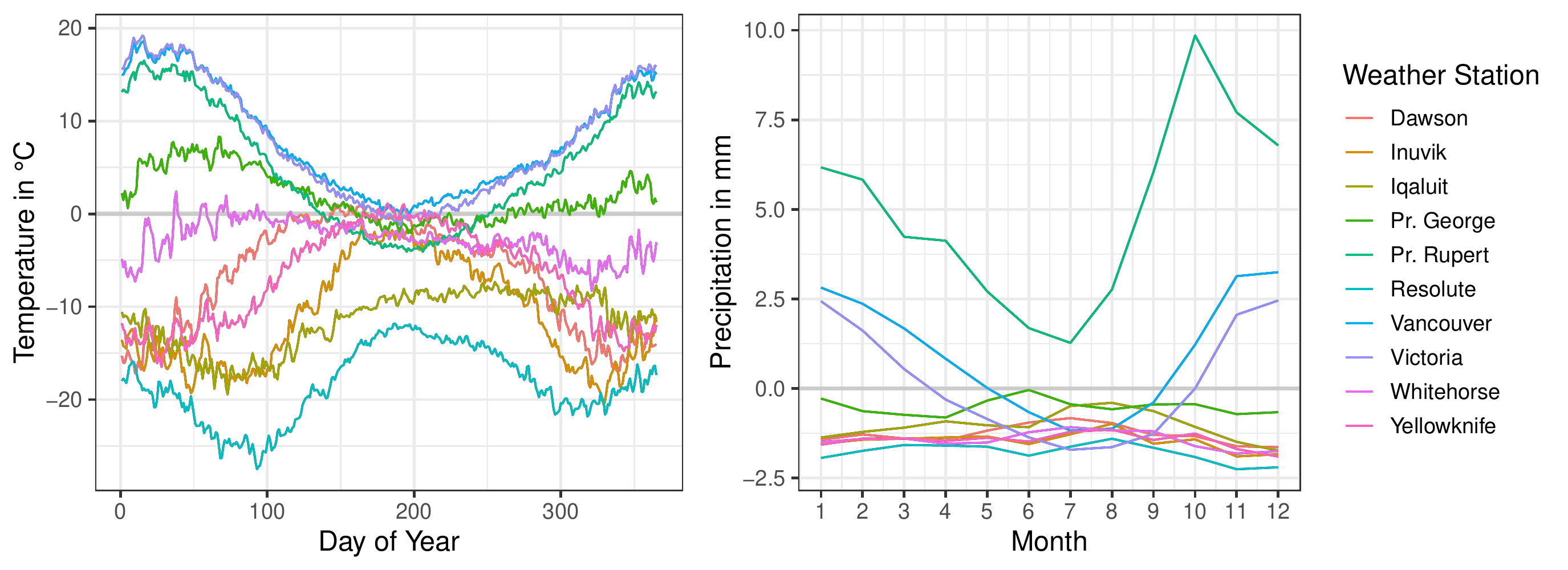}
\caption{Demeaned versions of the ten \code{canadianWeather} observations shown in Figure~\ref{fig:ggplot}. The curves have been obtained by \code{canadWeather - meanFunction(canadWeather)}, i.e., the bivariate mean function of all $35$ weather stations has been subtracted from each observation. The horizontal gray lines mark zero, corresponding to the original mean function.}
\label{fig:plotArith}
\end{figure}

With the \pkg{funData} package, mathematical operations can directly be applied to functional data objects, with the calculation made pointwise and the return being again an object of the same class. The operations  build on the \code{Arith} and \code{Math} \textit{group generics} for S4 classes.
We can for example convert the Canadian temperature data from Celsius to Fahrenheit:
\begin{Schunk}
\begin{Sinput}
R> 9/5 * dailyTemp + 32
\end{Sinput}
\end{Schunk}
or calculate the logarithms of the CD4 count data:
\begin{Schunk}
\begin{Sinput}
R> log(cd4Counts)
\end{Sinput}
\end{Schunk}
Arithmetic operations such as sums or products are implemented for scalars and functional data objects as well as for two functional data objects. Note that in the last case, the functional data objects must have the same number of observations (in this case, the calculation is done with the $i$-th function of the first object and the $i$-th  function of the second object) or one object may have only one observation (in this case, the calculation is made with each function of the other object). This is particularly useful e.g.,\ for subtracting a common mean from all functions in an object, as in the following example, which uses the \code{meanFunction} method:
\begin{Schunk}
\begin{Sinput}
R> canadWeather - meanFunction(canadWeather)
\end{Sinput}
\end{Schunk}
Some of the demeaned curves are shown in Figure~\ref{fig:plotArith}. 
 Note that the functions also need to have the same observation points, which is especially important for \code{irregFunData} objects.

\begin{figure}
\begin{center}
\includegraphics[width = 0.7\textwidth]{\PathSoft/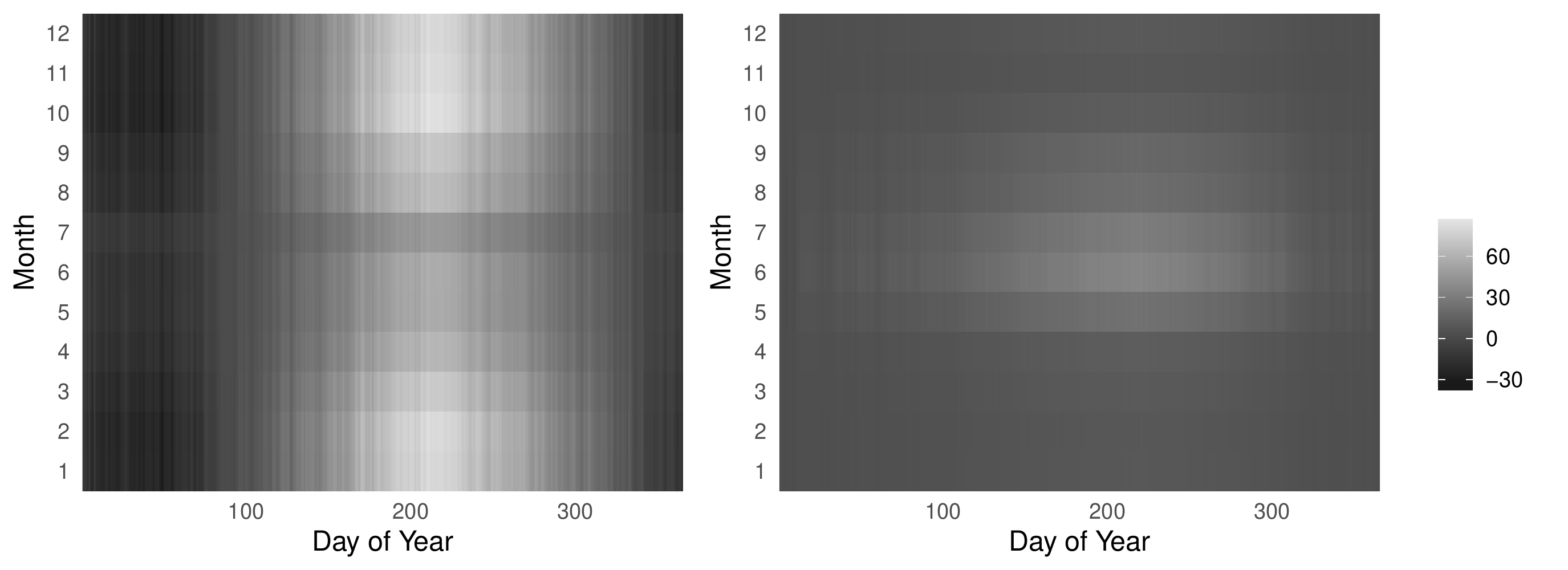}
\end{center}
\caption{Two observations of the tensor product of daily temperature and monthly precipitation from the Canadian weather data, calculated via \code{tensorProduct(dailyTemp, monthlyPrec)}. As seen in the plot, the domains of the functions have to be one-dimensional, but can be different. The result is an object of class \code{funData} on the two-dimensional domain $[1,356] \times [1,12]$ with $35^2 = 1225$ observations, from which two are shown here.}
\label{fig:tensorProduct}
\end{figure}

The \code{tensorProduct} function allows to calculate tensor products of functional data objects $f_1, f_2$ on one-dimensional domains $\calT_1, \calT_2$, respectively, i.e.,
\[f_\text{Tens}(t_1,t_2) = f_1(t_1) f_2(t_2) \quad t_1 \in \calT_1, t_2 \in \calT_2.\]
The following code calculates the tensor product of the Canadian weather data and the output shows that the result is a \code{funData} object on a two-dimensional domain:
\begin{Schunk}
\begin{Sinput}
R> tensorData <- tensorProduct(dailyTemp, monthlyPrec)
R> tensorData
\end{Sinput}
\begin{Soutput}
Functional data with 1225 observations of 2-dimensional support
argvals:
	1 2 ... 365		(365 sampling points)
	1 2 ... 12		(12 sampling points)
X:
	array of size 1225 x 365 x 12 
\end{Soutput}
\end{Schunk}
Two observations in \code{tensorData} are shown in Figure~\ref{fig:tensorProduct}. Note that for image data, a single observation has to be specified for plotting.

Another important aspect when working with functional data is integration, e.g.,\ in the context of principal component analysis or regression, where scalar products between functions replace the usual scalar products between vectors from multivariate analysis. The \pkg{funData} package implements two quadrature rules, \code{"midpoint"} and \code{"trapezoidal"} (the default). The data is always integrated over the full domain and in the case of multivariate functional data, the integrals are calculated for each element and the results are added. 
 For irregular data, the integral can be calculated on the observed points or they can be extrapolated linearly to the full domain. For the latter, curves with only one observation point are assumed to be constant.
 
Based on integrals, one defines the usual scalar product on $L^2(\calT)$
$ \scal{f}{g} = \int_\calT f(t) g(t) \drm t $
and the induced norm $\norm{f} = \scal{f}{f}^{1/2}$ for $f,g \in L^2(\calT)$. For multivariate functional data on domains $\calT_1 \times \ldots \times \calT_p$, the scalar product can be extended to
\[\myscal{f}{g} = \sum_{j = 1}^p f\h{j}(t) g\h{j}(t) \drm t\]
with the induced norm $\mynorm{f} = \myscal{f}{f}^{1/2}$. The multivariate scalar product can further be generalized by introducing weights $w_j > 0$ for each element \citep[cf.][]{HappGreven:2016}:
\begin{equation}
\myscal{f}{g}_w = \sum_{j = 1}^p w_j \scal{f\h{j}}{g\h{j}}.
\label{eq:weightedScalarProduct}
\end{equation}
Scalar products and norms are implemented for all three classes in the \pkg{funData} package. Here also, the scalar product can be calculated for pairs of functions $f_1 \usw f_N$ and $g_1 \usw g_N$, hence $\scal{f_i}{g_i}$, or for a sample $f_1 \usw f_N$ and a single function $g$, returning $\scal{f_i}{g}$. The \code{norm} function accepts some additional arguments, such as \code{squared} (logical, should the squared norm be calculated) or \code{weight} (a vector containing the weights $w_1 \usw w_p$ for multivariate functional data):
\begin{Schunk}
\begin{Sinput}
R> all.equal(scalarProduct(dailyTemp, dailyTemp), 
+            norm(dailyTemp, squared = TRUE))
\end{Sinput}
\begin{Soutput}
[1] TRUE
\end{Soutput}
\end{Schunk}

\subsection{Simulation toolbox}
\label{sec:simTools}
The \pkg{funData} package comes with a full simulation toolbox for univariate and multivariate functional data, which is a very useful feature when implementing and testing new methodological developments. The data is simulated based on a truncated Karhunen-Lo\`{e}ve representation of functional data, as for example in the simulation studies in \citet{ScheiplGreven:2016} or \citet{HappGreven:2016}. All examples in the following text use \code{set.seed(1)} before calling a simulation function for reasons of reproducibility. 

For univariate functions $x_i \colon \calT \to \IR$, the Karhunen-Lo\`{e}ve representation of a function $x_i$ truncated at $M \in \IN$ is given by
\begin{equation}
x_i(t) = \mu(t) + \sum_{m = 1}^M \xi_{i,m} \phi_m(t),\quad i = 1\usw N,~ t \in \calT
\label{eq:KarhunenUniv}
\end{equation}
with a common mean function $\mu(t)$ and principal component functions $\phi_m,~ m = 1 \usw M$.  The individual functional principal component scores $\xi_{i,m} = \scal{x_i}{ \phi_m}$ are realizations of random variables $\xi_m$ with $\IE(\xi_{m}) = 0$ and $\Var(\xi_{m}) = \lambda_m$ with eigenvalues $\lambda_m \geq 0$ that decrease towards $0$. This representation is valid for domains of arbitrary dimension, hence also for $\calT \subset \IR^2$ (images) or $\calT \subset \IR^3$ (3D images).

The simulation algorithm constructs new data from a system of $M$ orthonormal eigenfunctions $\phi_1 \ldots \phi_M$ and scores $\xi_{i,m}$ according to the Karhunen-Lo\`{e}ve representation Equation~\ref{eq:KarhunenUniv} with $\mu(t) \equiv 0$. For the eigenfunctions, the package implements Legendre polynomials, Fourier basis functions and eigenfunctions of the Wiener process including some variations (e.g.,\ Fourier functions plus an orthogonalized version of the linear function). The scores are generated via
\begin{equation}
 \xi_{i,m} \iidsim \N(0, \lambda_m),~ m = 1 \usw M,~ i = 1 \usw N.
 \label{eq:simScores}
\end{equation}
For the eigenvalues $\lambda_m$, one can choose between a  linear ($\lambda_m = \frac{M-m+1}{M}$) or exponential decrease ($\exp(-\tfrac{m+1}{2})$) or those of the Wiener process. New eigenfunctions and eigenvalues can be added to the package in an easy and modular manner.

\begin{figure}
\begin{center}
\includegraphics[width = 0.7\textwidth]{\PathSoft/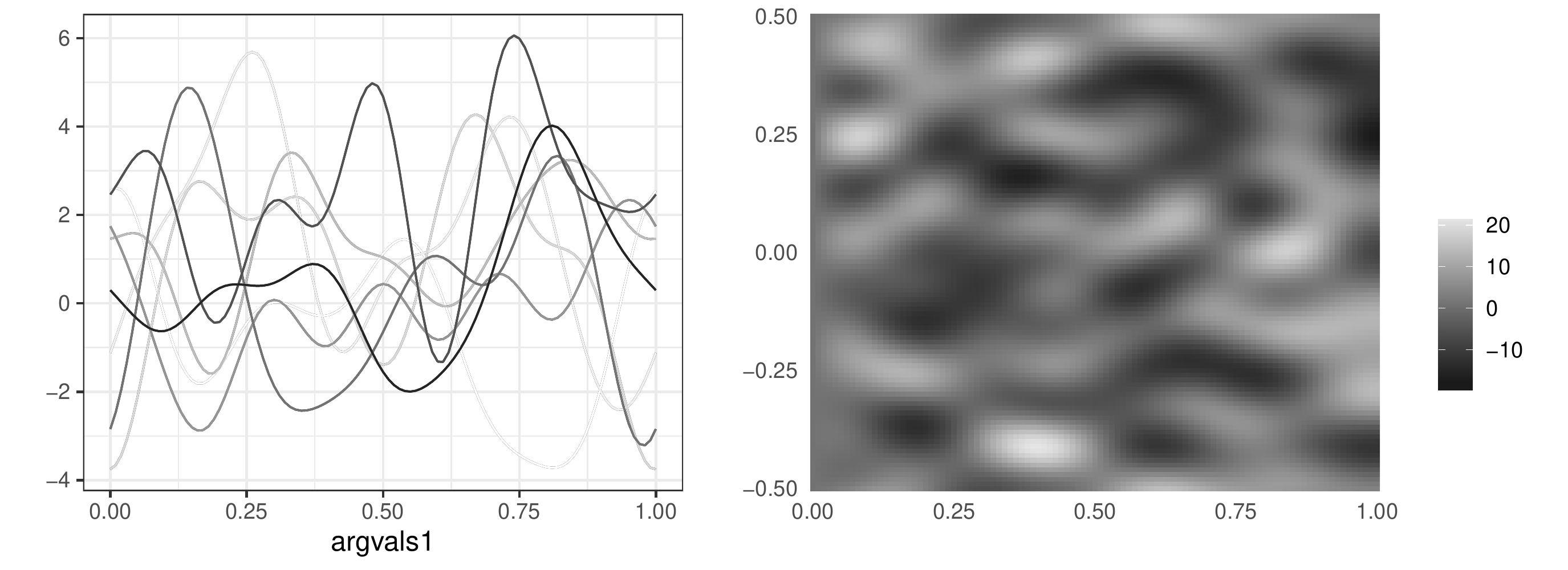}
\end{center}
\caption{Left: $N =8$ simulated curves on $[0,1]$ based on the first $M = 10$ Fourier basis functions and eigenvalues with a linear decrease. Right: One simulated image on $[0,1] \times[-0.5,0.5]$ based on tensor products of $M_1 = 10$ eigenfunctions of the Wiener process on $[0,1]$ and $M_2 = 12$ Fourier basis functions on $[-0.5,0.5]$ and linearly decreasing eigenvalues.}
\label{fig:simUniv}
\end{figure}

The next code chunk simulates $N =8$ curves on the one-dimensional observation grid $\{0, 0.01, 0.02 \usw 1\}$ based on the first $M = 10$ Fourier basis functions on $[0,1]$ and eigenvalues with a linear decrease:
\begin{Schunk}
\begin{Sinput}
R> simUniv1D <- simFunData(N = 8, argvals = seq(0,1,0.01),
+                          eFunType = "Fourier", eValType = "linear", M = 10)
\end{Sinput}
\end{Schunk}
The function returns a list with $3$ entries: The simulated data (\code{simData},  a \code{funData} object shown in Figure~\ref{fig:simUniv}),  the true eigenvalues (\code{trueVals}) and eigenfunctions (\code{trueFuns}, also as a \code{funData} object).

For simulating functional data on a two- or higher dimensional domain, \code{simFunData} constructs eigenfunctions based on tensor products of univariate eigenfunctions. The user thus has to supply the parameters that relate to the eigenfunctions as a list (for \code{argvals}) or as a vector (\code{M} and \code{eFunType}), containing the information for each marginal. The total number of eigenfunctions equals to the product of the entries of $M$. The following example code  simulates $N = 5$ functions on $\calT = [0,1] \times[-0.5,0.5]$. The eigenfunctions are calculated as tensor products of $M_1 = 10$ eigenfunctions of the Wiener process on $[0,1]$ and $M_2 = 12$  Fourier basis functions on $[-0.5,0.5]$. In total, this leads to $M = M_1 \cdot M_2 = 120$ eigenfunctions. For each eigenfunction and each observed curve, the scores $\xi_{i,m}$ are generated as in Equation~\ref{eq:simScores} with linearly decreasing eigenvalues:
\begin{Schunk}
\begin{Sinput}
R> argvalsList <- list(seq(0,1,0.01), seq(-0.5,0.5, 0.01))
R> simUniv2D <- simFunData(N = 5, argvals = argvalsList,
+       eFunType = c("Wiener", "Fourier"), eValType = "linear", M = c(10,12))
\end{Sinput}
\end{Schunk}
 The first simulated image is shown in Figure~\ref{fig:simUniv}. As for functions on one-dimensional domains, the function returns the simulated data together with the true eigenvalues and eigenfunctions. 

For multivariate functional data, the simulation is based on the multivariate version of the Karhunen-Lo\`{e}ve Theorem \citep{HappGreven:2016} for multivariate functional data $x_i = (x_i\h{1} \usw x_i\h{p})$ truncated at $M \in \IN$
\begin{equation}
x_i(t) = \mu (t) + \sum_{m = 1}^M \rho_{i,m} \psi_m(t),\quad i = 1\usw N,~ t = (t_1 \usw t_p) \in \calT_1 \times \ldots \times \calT_p
\label{eq:KarhunenMultiv}
\end{equation}
with a multivariate mean function $\mu$ and multivariate eigenfunctions $\psi_m$ that have the same structure as $x_i$ (i.e., if $x_i$  consists of a function and an image, then $\mu$ and $\psi_m$ will also be bivariate functions, consisting of a function and an image). 
The individual scores $\rho_{i,m} = \myscal{x_i}{\psi_m}$ for each observation $x_i$ and each eigenfunction $\psi_m$ are real numbers and have the same properties as in the univariate case, i.e., they are realizations of random variables $\rho_m$ with $\IE(\rho_{m}) = 0$ and $\Var(\rho_{m}) = \nu_m$ with eigenvalues $\nu_m \geq 0$ that again form a decreasing sequence that converges towards $0$.
As in the univariate case, the multivariate functions are simulated based on eigenfunctions and scores according to Equation~\ref{eq:KarhunenMultiv} with $\mu(t) \equiv 0$. The scores are sampled independently from a $\N(0, \nu_m)$ distribution with decreasing eigenvalues $\nu_m$, analogously to Equation~\ref{eq:simScores}.
For the construction of multivariate eigenfunctions, \citet{HappGreven:2016} propose two approaches based on univariate orthonormal systems, which are both implemented in the \code{simMultiFunData} function.


Calling \code{simMultiFunData} with the option \code{"split"} constructs multivariate eigenfunctions by splitting orthonormal functions into $p$ pieces and shifting them to where the elements should be defined. This works only for functions on one-dimensional domains. 
The following code simulates $N = 7$ bivariate functions on $[-0.5, 0.5]$ and $[0,1]$, based on $M = 10$ Fourier basis functions and linearly decreasing eigenvalues.
\begin{Schunk}
\begin{Sinput}
R> argvalsList <- list(seq(-0.5,0.5,0.01), seq(0,1,0.01))
R> simMultiSplit <- simMultiFunData(N = 7, argvals = argvalsList, 
+                      eFunType = "Fourier", eValType = "linear", M = 10,
+                      type = "split")
\end{Sinput}
\end{Schunk}

\begin{figure}
\begin{center}
\includegraphics[width = 0.7\textwidth]{\PathSoft/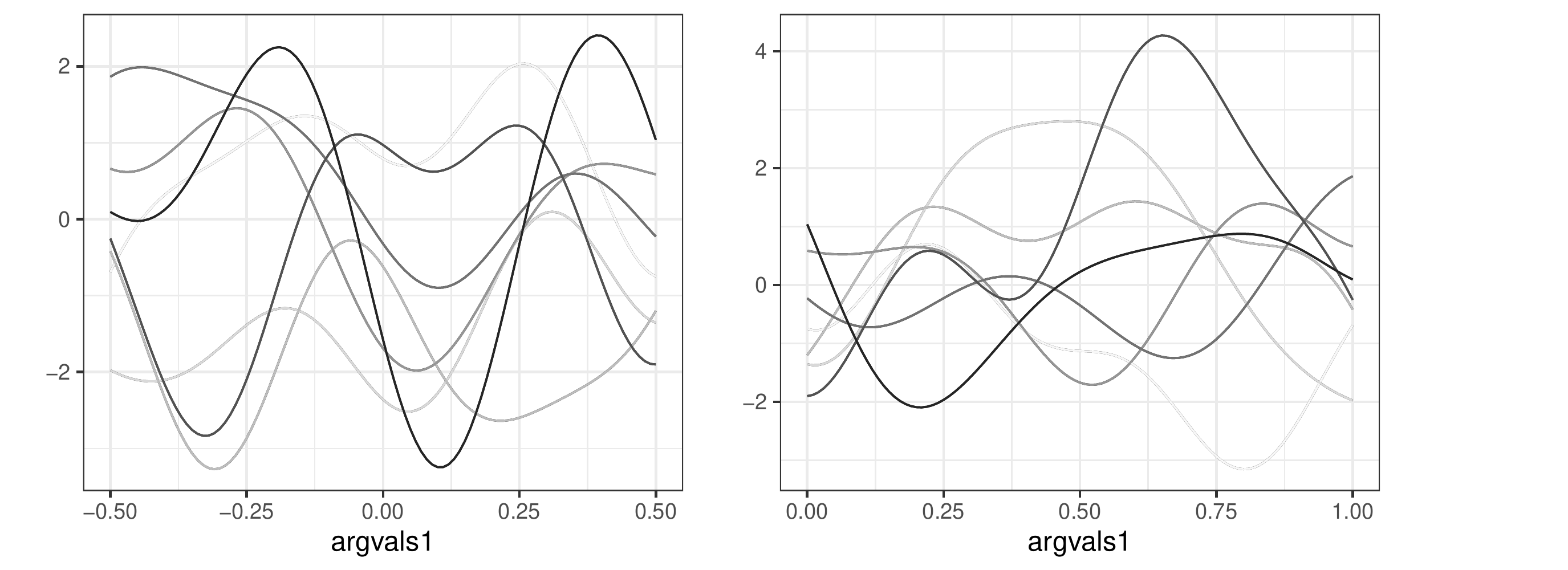}
\end{center}
\caption{$N =7$ simulated bivariate curves on $[-0.5, 0.5]$ and $[0,1]$ with eigenfunctions obtained from  the first $M = 10$ Fourier basis functions by the splitting algorithm (\code{type = "split"}) and linearly decreasing eigenvalues.}
\label{fig:simMultiv_split}
\end{figure}

\begin{figure}
\begin{center}
\includegraphics[width = 0.7\textwidth]{\PathSoft/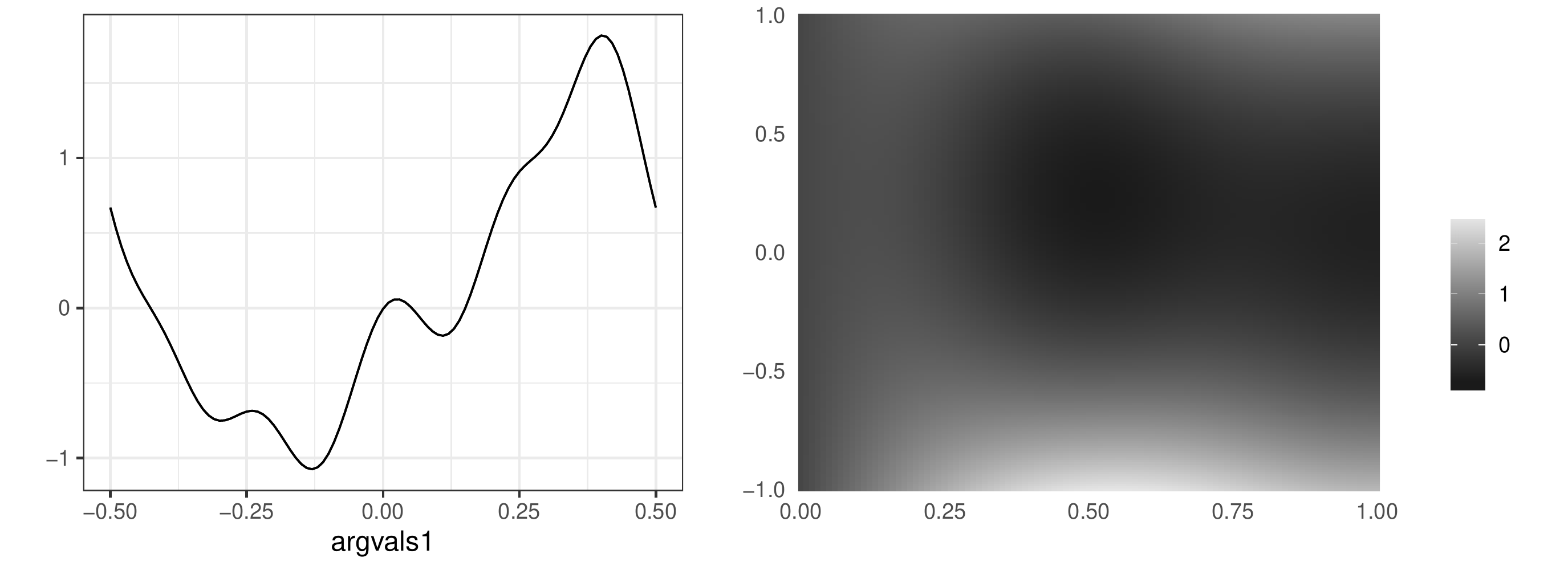}
\end{center}
\caption{One observation of simulated bivariate data on $[-0.5, 0.5]$ and $[0,1] \times [-1,1]$ using weighted orthonormal elements (\code{type = "weighted"}). See text for details.}
\label{fig:simMultiv_weight}
\end{figure}

As an alternative, multivariate eigenfunctions can be constructed as weighted versions of univariate eigenfunctions. With this approach, one can also simulate multivariate functional data on different dimensional domains, e.g.,\ functions and images. It is implemented in \pkg{funData}'s \code{simMultiFunData} method using the option \code{type = "weighted"}.
The following code simulates $N = 5$ bivariate functions on $\calT_1 = [-0.5, 0.5]$ and $\calT_2 = [0,1] \times [-1,1]$. The first elements of the eigenfunctions are derived from $M_1 = 12$ Fourier basis functions on $\calT_1$ and the second elements of the eigenfunctions are constructed from tensor products of $4$ eigenfunctions of the Wiener process on $[0,1]$ and $3$ Legendre polynomials on $[-1,1]$, which give together $M_2 = 12$ eigenfunctions on $\calT_2$. The scores are sampled using exponentially decreasing eigenvalues:
\begin{Schunk}
\begin{Sinput}
R> argvalsList <- list(seq(-0.5,0.5,0.01), list(seq(0,1,0.01), seq(-1,1,0.01)))
R> simMultiWeight <- simMultiFunData(N = 5, argvals = argvalsList,
+                       eFunType = list("Fourier", c("Wiener", "Poly")),
+                       eValType = "exponential",  M = list(12, c(4,3)), 
+                       type = "weighted")
\end{Sinput}
\end{Schunk}
In both cases, the result contains the simulated data as well as the eigenfunctions and eigenvalues.
The simulated functions are shown in Figures~\ref{fig:simMultiv_split} and~\ref{fig:simMultiv_weight}. For more technical details on the construction of the eigenfunctions, see \citet{HappGreven:2016}.

\begin{figure}
\begin{center}
\includegraphics[width = 0.7\textwidth]{\PathSoft/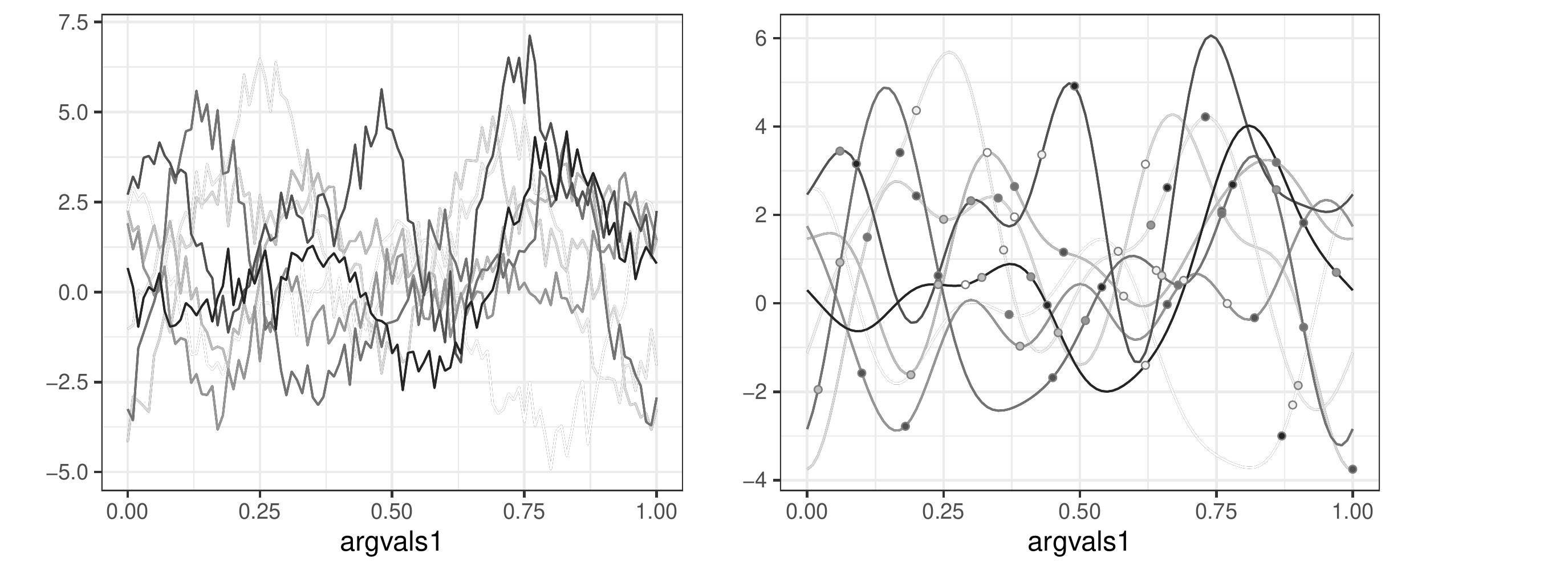}
\end{center}
\caption{%
Transforming the simulated univariate functions in \code{simUniv1D} (see Figure~\ref{fig:simUniv}). Left: Adding noise with a standard deviation of $\sigma = 0.5$. Right: The effect of sparsification, keeping five to ten observations per curve. Solid lines show the original data, filled dots correspond to the observed values of the sparsified version.
}
\label{fig:simUniv_Err}
\end{figure}

\begin{figure}
\begin{center}
\includegraphics[width = 0.7\textwidth]{\PathSoft/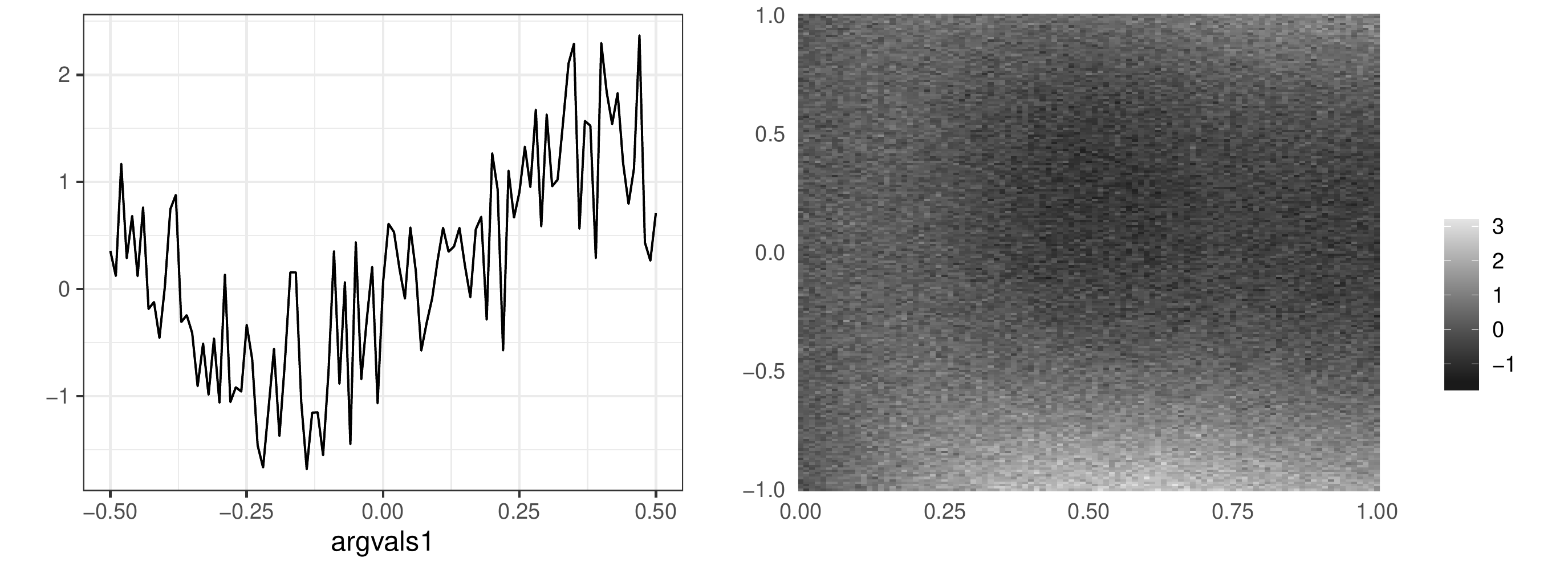}

\includegraphics[width = 0.7\textwidth]{\PathSoft/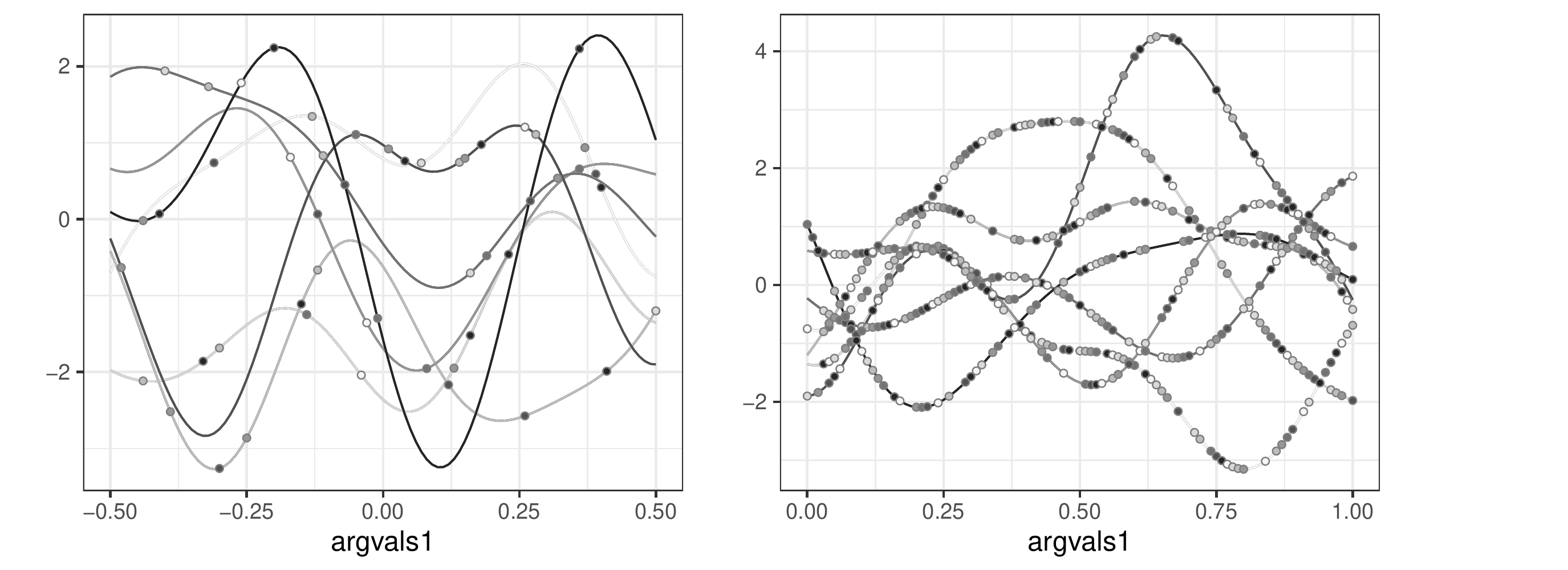}
\end{center}
\caption{Transforming the simulated bivariate data. First row: A noisy version of the first observation of \code{simMultiWeight} (see Figure~\ref{fig:simMultiv_weight}). Second row: All $7$ observations of \code{simMultiSplit} after sparsification (see Figure~\ref{fig:simMultiv_split}). Solid lines show the original data, filled dots correspond to the observed values of the sparsified version. Note that the standard deviation in the noise as well as the degree of sparsification varies across elements.}
\label{fig:simMultiv_Err}
\end{figure}

Once simulated, the data can be further processed by adding noise (function \code{addError}) or by artificially deleting measurements (sparsification, function \code{sparsify}). The latter is done in analogy to \citet{YaoEtAl:2005}. Examples for modified versions of simulated functions can be computed as follows:
\begin{Schunk}
\begin{Sinput}
R> addError(simUniv1D$simData, sd = 0.5)
R> sparsify(simUniv1D$simData, minObs = 5, maxObs = 10)
R> 
R> addError(simMultiWeight$simData, sd = c(0.5, 0.3))
R> sparsify(simMultiSplit$simData, minObs = c(5, 50), maxObs = c(10, 80))
\end{Sinput}
\end{Schunk}
The results are shown in Figure~\ref{fig:simUniv_Err} for the univariate case and in Figure~\ref{fig:simMultiv_Err} for the multivariate case.

\FloatBarrier

\section[The MFPCA package]{The \pkg{MFPCA} package}
\label{sec:MFPCApack}

The \pkg{MFPCA} package implements multivariate functional principal component analysis (MFPCA) for data on potentially different dimensional domains \citep{HappGreven:2016}.\footnote{Not to be confused with \textit{multilevel} functional principal component analysis, which is implemented in \pkg{refund} as \code{mfpca}.} It  heavily builds upon the \pkg{funData} package, i.e., all functions are implemented as functional data objects. The \pkg{MFPCA} package thus illustrates the use of \pkg{funData} as a universal basis for implementing new methods for functional data.
Section~\ref{sec:MFPCAtheory} gives a short review of the MFPCA methodology and  Section~\ref{sec:MFPCAimplement} describes the implementation including a detailed description of the main functions and a practical case study. For theoretical details, please refer to \citet{HappGreven:2016}.

\subsection{Methodological background}
\label{sec:MFPCAtheory}

The basic idea of MFPCA is to extend functional principal component analysis to multivariate functional data on different dimensional domains. The data is assumed to be iid samples $x_1 \usw x_N$  of a random process $X = (X\h{1} \usw X\h{p})$ with $p$ elements $X\h{j} \in L^2(\calT_j)$ on domains $\calT_j \subset \IR^{d_j}$ with potentially different dimensional dimensions $d_j \in \IN$. \citet{HappGreven:2016} provide an algorithm to estimate multivariate functional principal components and scores based on their univariate counterparts.
The algorithm starts with demeaned samples $x_1 \usw x_N$ and consists of four steps:
\begin{enumerate}
\item \label{algo:univDecomp}
Calculate a univariate functional principal component analysis for each element $ j = 1\usw p$. This results in principal component functions  $\hat \phi_1\h{j} \usw \hat \phi_{M_j}\h{j}$ and principal component scores $\hat \xi_{i,1}\h{j} \usw \hat \xi_{i, M_j}\h{j}$ for each observation unit $i = 1 \usw N$ and suitably chosen truncation lags $M_j$.
\item \label{algo:combCovMat}
Combine all coefficients into one big matrix $\Xi \in \IR^{N \times M_+}$ with $M_+ = M_1 + \ldots + M_p$, having rows
\[
\Xi_{i, \cdot} = \left(\hat \xi_{i,1}\h{1} \usw \hat \xi_{i, M_1}\h{1} \usw \hat \xi_{i,1}\h{p} \usw \hat \xi_{i, M_p}\h{p} \right)
\]
and estimate the joint covariance matrix $\hat Z = \frac{1}{N-1} \Xi\trans \Xi$.
\item \label{algo:eigenCombCovMat}
Find eigenvectors $\hat c_m$ and eigenvalues $\hat \nu_m$ of $\hat Z$ for $m = 1 \usw M$ for some truncation lag $M \leq M_+$.
\item Calculate estimated multivariate principal component functions $\hat \psi_m$ and scores $\hat \rho_{i,m}$ based on the results from steps~\ref{algo:univDecomp} and~\ref{algo:eigenCombCovMat}:
\[
 \hat \psi_m\h{j} = \sum_{n = 1}^{M_j} [\hat c_m]_n\h{j} \hat \phi_n\h{j}, \qquad  
  \hat \rho_{i,m} = \sum_{j = 1}^p \sum_{n=1}^{M_j} [\hat c_m]_n\h{j} \hat \xi_{i,n}\h{j} =  \Xi_{i, \cdot}  \hat c_m,\quad m = 1\usw M.
\]

\end{enumerate}
The advantage of MFPCA with respect to univariate FPCA for each component can be seen in steps~\ref{algo:combCovMat} and~\ref{algo:eigenCombCovMat}: The multivariate version takes covariation between the different elements into account, by using the joint covariance of the scores of all elements. 

As discussed for the simulation in Section~\ref{sec:simTools}, the multivariate principal component functions will have the same structure as the original samples, i.e., $\hat \psi_m = \left( \hat \psi_m\h{1} \usw \hat \psi_m\h{p} \right)$ with $\hat \psi_m\h{j} \in L^2(\calT_j)$ for $m = 1 \usw M$. The scores $\hat \rho_{i,m}$ give the individual weight of each observation $x_i$ for the principal component $\hat \psi_m$ in the empirical version of the truncated multivariate Karhunen-Lo\`eve representation:
\begin{equation}
x_i \approx \hat \mu + \sum_{m = 1}^M \hat \rho_{i,m} \hat \psi_m,
\label{eq:empKL}
\end{equation}
with $\hat \mu$ being an estimate for the multivariate mean function, cf. Equation~\ref{eq:KarhunenMultiv}.

In some cases, it might be of interest to replace the univariate functional principal component analysis in step~\ref{algo:univDecomp} by a representation in terms of fixed basis functions $B_1\h{j} \usw B_{K_j}\h{j}$, such as splines. In \citet{HappGreven:2016} it is shown how the algorithm can be extended to arbitrary basis functions in $L^2(\calT_j)$. Mixed approaches with some elements expanded in principal components and others for instance in splines are also possible.
Another very likely case is that the elements of the multivariate functional data differ in their domain, range or variation. For this case, \citet{HappGreven:2016} develop a weighted version of MFPCA with weights $w_j > 0$ for the different elements $j = 1 \usw p$. The weights have to be chosen depending on the data and the question of interest. One possible choice is to use the inverse of the integrated pointwise variance for the weights, as  proposed in \citet{HappGreven:2016}: $w_j = \left( \int_{\calT_j} \widehat{\Var}(X\h{j}(t))\drm t \right)\inv $.

\subsection{MFPCA implementation}
\label{sec:MFPCAimplement}

The main function in the \pkg{MFPCA} package is \code{MFPCA}, that calculates the multivariate functional principal component analysis. It requires as input arguments a \code{multiFunData} object for which the MFPCA should be calculated, the number of principal components \code{M} to calculate and a list \code{uniExpansions} specifying the univariate representations to use in step~\ref{algo:univDecomp}. It returns an object of class \code{MFPCAfit}, which has methods for printing, plotting and summarizing. Before discussing the detailed options, we illustrate the usage of \code{MFPCA} with a real data application.

\subsubsection{Case study: Calculating the MFPCA for the Canadian weather data}
The following example calculates a multivariate functional principal component analysis for the bivariate Canadian weather data with three principal components, using  univariate FPCA with five principal components for the daily temperature (element 1) and univariate FPCA with four principal components for the monthly precipitation (element 2). The univariate expansions are specified in a list with two list entries (one for each element) and are then passed to the main function:
\begin{Schunk}
\begin{Sinput}
R> uniExpansions <- list(list(type = "uFPCA", npc = 5), # temperature element
+                      list(type = "uFPCA", npc = 4)) # precipitation element
R> MFPCAweather <- MFPCA(canadWeather, M = 3, uniExpansions = uniExpansions)
\end{Sinput}
\end{Schunk}
The full analysis takes roughly nine seconds on a standard laptop, with most time spent for the univariate decompositions (if the elements are for example expanded in penalized splines, the total calculation time reduces to one second).

The resulting object \code{MFPCAweather} contains the following elements: the multivariate mean function (\code{meanFunction}, as the data is demeaned automatically before the analysis), the empirical multivariate principal component functions (\code{functions}), the individual scores for each city (\code{scores}) and the estimated eigenvalues (\code{values}). Two additional elements can be used for calculating out-of-sample predictions (\code{vectors} and \code{normFactors}). The \code{summary} function gives a basic overview of the results.
\begin{Schunk}
\begin{Sinput}
R> summary(MFPCAweather)
\end{Sinput}
\begin{Soutput}
3 multivariate functional principal components estimated with 2 elements, each.
                     * * * * * * * * * *                     
                                     PC 1     PC 2     PC 3
Eigenvalue                       1.55e+04 1.48e+03 3.30e+02
Proportion of variance explained 8.96e-01 8.54e-02 1.90e-02
Cumulative proportion            8.96e-01 9.81e-01 1.00e+00
\end{Soutput}
\end{Schunk}
The eigenvalues here are rapidly decreasing, i.e., the first principal component already explains almost $90\%$ of the variability in the data. The decrease of the eigenvalues is graphically illustrated by the \code{screeplot} function (see Figure~\ref{fig:MFPCAscreeplot} in the appendix).

\begin{figure}[ht]
\centering
\includegraphics[width = 0.7\textwidth]{\PathSoft/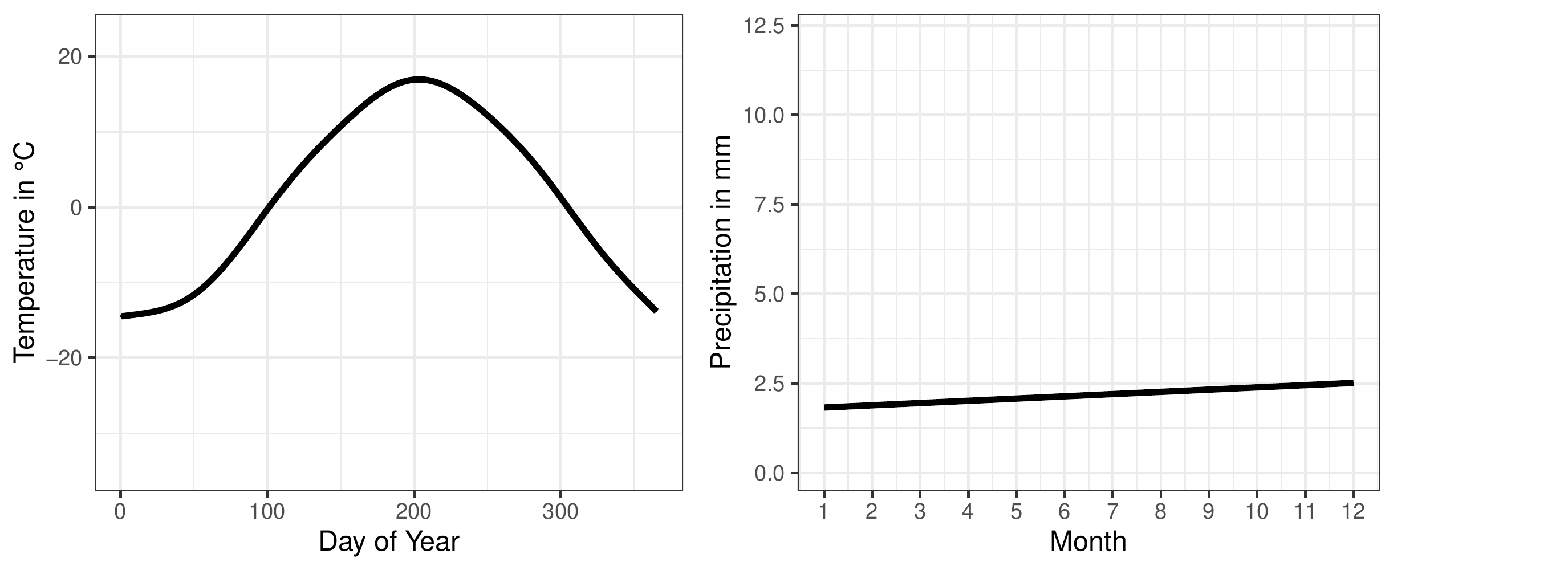}

\includegraphics[width = 0.7\textwidth]{\PathSoft/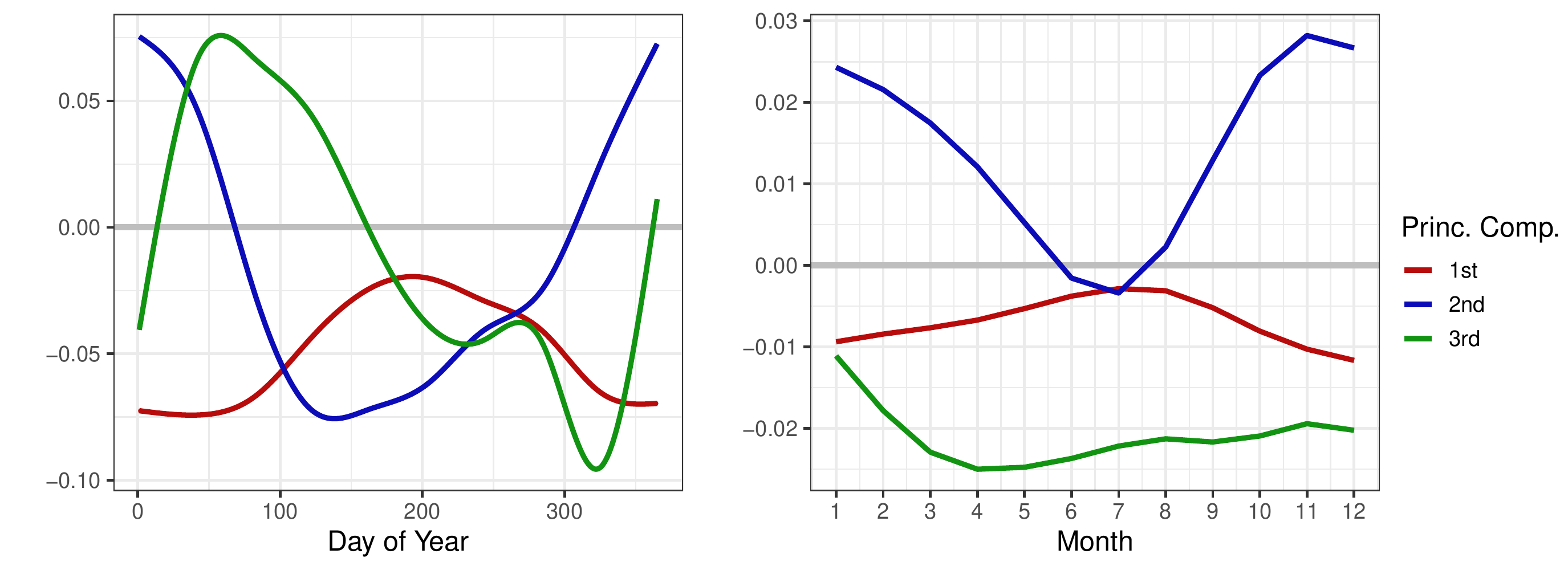}
\caption{MFPCA results for the Canadian weather data. First row: The bivariate mean function, which is subtracted from the data before calculating the MFPCA. Second row: The first three bivariate functional principal components. The gray horizontal lines in the principal component plots mark zero.}
\label{fig:MFPCA_funs}
\end{figure}

All functions in \code{MFPCAweather} are represented as functional data objects and can thus be plotted using the methods provided by the \pkg{funData} package (see Figure~\ref{fig:MFPCA_funs}). The mean function of the temperature element is seen to have low values below -10°C in the winter and a peak at around 15°C in the summer, while the mean of the monthly precipitation data is slightly increasing over the year. The first principal component is negative for both elements, i.e., weather stations with positive scores will in general have lower temperatures and less precipitation than on average. The difference is more pronounced in the winter than in the summer, as both the temperature as well as the precipitation element of the first principal component have more negative values in the winter period. This indicates that there is covariation between both elements, that can be captured by the MFPCA approach. An alternative visualization, plotting the principal component as perturbation of the mean as in the \pkg{fda} package, can be obtained via \code{plot(MFPCAweather)} (see Figure~\ref{fig:plotMFPCAfit} in the appendix).
In total, the first bivariate eigenfunction can be associated with arctic and continental climate, characterized by low temperatures, especially in the winter, and less precipitation than on average. 
Weather stations with negative score values will show an opposite behavior, with higher temperatures and more rainfall than on average, particularly in the winter months. This is typical for maritime climate. 

\begin{figure}[ht]
\centering
\includegraphics[width = 0.7\textwidth]{\PathSoft/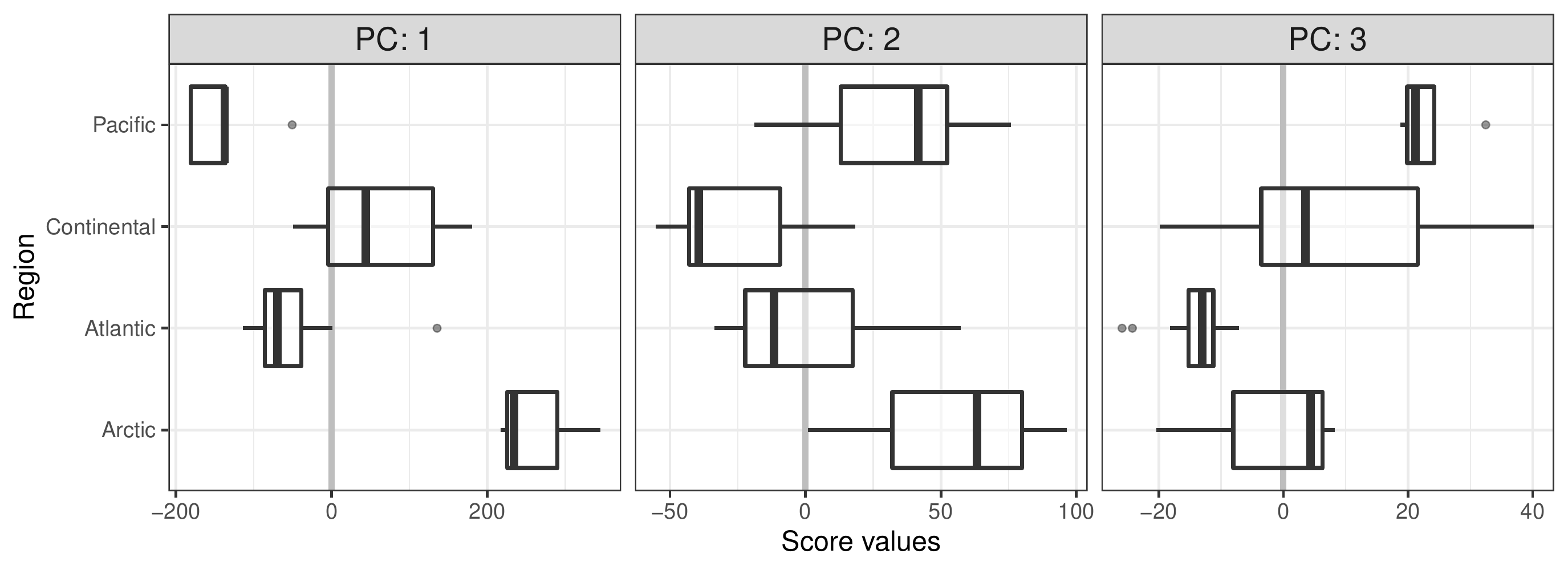}
\caption{Scores of the first three bivariate functional principal components (PCs) for the  Canadian weather data depending on the region of each weather station. The gray vertical lines mark zero.}
\label{fig:MFPCA_scores}
\end{figure}

The estimated scores for the first principal component support this interpretation, as weather stations in arctic and continental areas mainly have positive scores, while stations in the coastal areas have negative values in most cases (see Figure~\ref{fig:MFPCA_scores}). Moreover, weather stations in the arctic and pacific regions are seen to have more extreme score values than those in continental areas and on the Atlantic coast, meaning that the latter have a more moderate climate. An alternative visualization of the scores is given by the \code{scoreplot} function (see Figure~\ref{fig:plotMFPCAscores} in the appendix).

\subsubsection[More details on the MFPCA function]{More details on the  \code{MFPCA} function}

In the above example, both univariate elements have been decomposed in univariate functional principal components in step~\ref{algo:univDecomp}. The \pkg{MFPCA} package implements some further options for the univariate expansions, that can easily be extended in a modular way. The most common basis expansions are \code{uFPCA} (univariate PCA) and \code{splines1D / splines1Dpen} (splines) for elements on a one-dimensional domain and \code{splines2D / splines2Dpen} (tensor splines) and  \code{DCT2D/DCT3D} (tensor cosine basis) for elements on higher dimensional domains. If data has been smoothed, for example in a preprocessing step, the basis functions and coefficients can also be passed using \code{type = "given"} for the univariate basis expansions. All currently implemented basis expansions are presented in detail in the appendix. 

\begin{figure}
\centering
\includegraphics[width = 0.7\textwidth]{\PathSoft/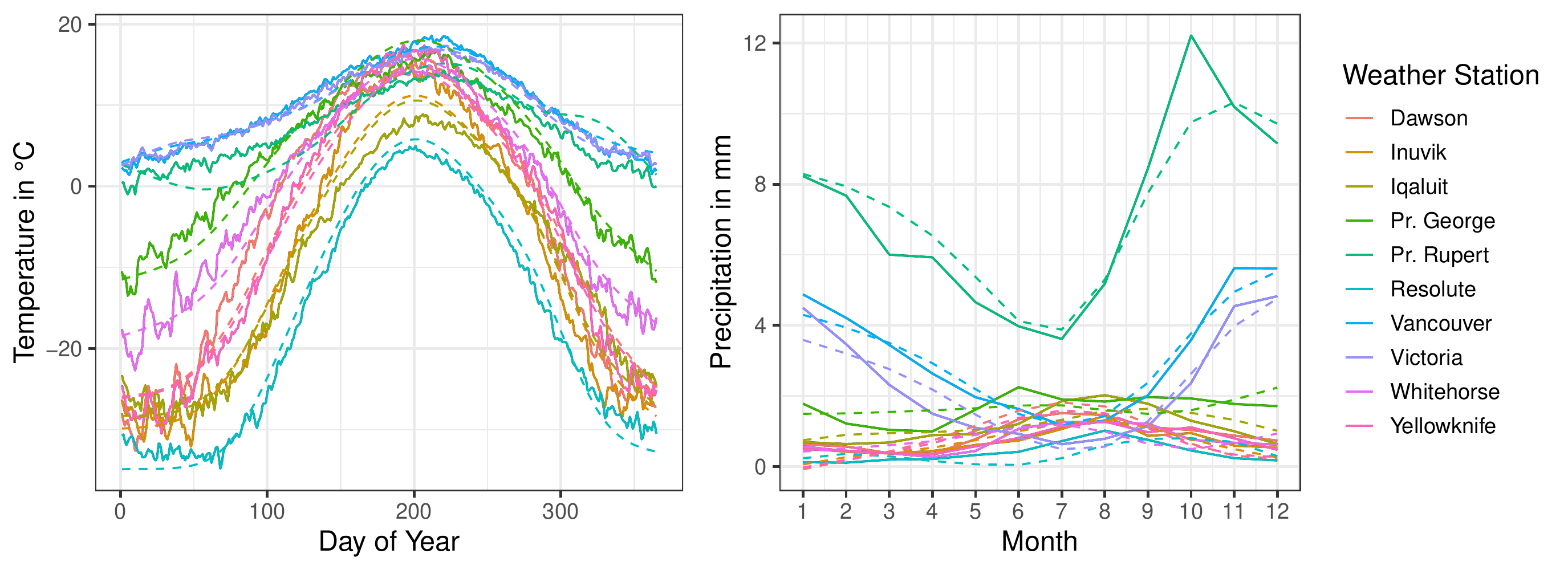}
\caption{The ten observations of the bivariate Canadian weather data shown in Figure~\ref{fig:ggplot} (solid lines) and their reconstruction (dashed lines) based on the truncated Karhunen-Lo\`{e}ve representation with estimates found by a weighted version of MFPCA (\code{MFPCAweatherFit}).}
\label{fig:MFPCA_fit}
\end{figure}

With the mean function, the principal components and the individual scores calculated in the \code{MFPCA} function, the observed functions $x_1 \usw x_N$  can be reconstructed based on the truncated Karhunen-Lo\`{e}ve representation with plugged-in estimators as in Equation~\ref{eq:empKL}.
The reconstructions can be obtained by setting the option \code{fit = TRUE}, which adds a multivariate functional data object \code{fit} with $N$ observations to the result object, where the $i$-th entry corresponds to the reconstruction $\hat x_i$ of an observation $x_i$. 
For a weighted version of MFPCA, the weights can be supplied to the \code{MFPCA} function in form of a vector \code{weights} of length $p$, containing the weights $w_j > 0$ for each element $j = 1 \usw p$. Both options are used in the following example for the \code{CanadWeather} data, which uses the weights based on the integrated pointwise variance, as discussed in \citet{HappGreven:2016}:
\begin{Schunk}
\begin{Sinput}
R> varTemp <- funData(argvals = canadWeather[[1]]@argvals, 
+                   X = matrix(apply(canadWeather[[1]]@X, 2, var), nrow = 1))
R> varPrec <- funData(argvals = canadWeather[[2]]@argvals, 
+                   X = matrix(apply(canadWeather[[2]]@X, 2, var), nrow = 1))
R> 
R> weightWeather <- c(1/integrate(varTemp), 1/integrate(varPrec))
\end{Sinput}
\end{Schunk}
Given the weights, the MFPCA is calculated including reconstructions of the observed functions:
\begin{Schunk}
\begin{Sinput}
R> MFPCAweatherFit <- MFPCA(canadWeather, M = 3, 
+                           uniExpansions = uniExpansions, 
+                           weights = weightWeather, fit = TRUE)
\end{Sinput}
\end{Schunk}
Figure~\ref{fig:MFPCA_fit} shows some original functions of the \code{canadWeather} data and their reconstructions saved in \code{MFPCAweatherFit}. Alternatively, reconstructions can be obtained by applying the \code{predict} function to the \code{MFPCAfit} object. 

If elements are expanded in fixed basis functions, the number of basis functions that are needed to represent the data well will in general be quite high, particularly for elements with higher dimensional domains. As a consequence, the covariance matrix of all scores in step~\ref{algo:combCovMat} of the estimation algorithm can become large and the eigendecompositions in step~\ref{algo:eigenCombCovMat} can get computationally very demanding. By setting the option \code{approx.eigen = TRUE}, the eigenproblem is solved approximately using the  augmented implicitly restarted Lanczos bidiagonalization algorithm  \citep[IRLBA,][]{BaglamaReichel:2005} implemented in the \pkg{irlba} package \citep{irlba}.
The \pkg{MFPCA} function also implements nonparametric bootstrap on the level of functions to quantify the uncertainty in the estimation  \citep[cf.][]{HappGreven:2016}. Setting \code{bootstrap = TRUE}  calculates pointwise bootstrap confidence bands for the principal component functions and bootstrap confidence bands for the associated eigenvalues.

\section{Summary and outlook}
\label{sec:summaryOut}

The \pkg{funData} package implements functional data in an object-oriented manner. The aim of the package is to provide a flexible and unified toolbox for dense univariate and multivariate functional data with different dimensional domains as well as irregular functional data. The package implements basic utilities for creating, accessing and modifying the data, upon which other packages can be built. This distinguishes the \pkg{funData} package from other packages for functional data, that either do not provide a specific data structure together with basic utilities or mix this aspect with the implementation of advanced methods for functional data.

The \pkg{funData} package implements three classes for representing functional data based on the observed values and without any further assumptions such as basis function representations. The classes follow a unified approach for representing and working with the data, which means that the same methods are implemented for all the three classes (polymorphism).  The package further includes a full simulation toolbox for univariate and multivariate functional data on one- and higher dimensional domains. This is a very useful feature when implementing and testing new methodological developments. 

The \pkg{MFPCA} package is an example for an advanced methodological package, which builds upon the \pkg{funData} functionalities. It implements a new approach, multivariate functional principal component analysis  for data on different dimensional domains \citep{HappGreven:2016}. All calculations relating to the functional data, data input and output use the basic \pkg{funData} classes and methods.

Both packages, \pkg{funData} and \pkg{MFPCA}, are publicly available on CRAN (\url{https://CRAN.R-project.org}) and GitHub (\url{https://github.com/ClaraHapp}). They come with a comprehensive documentation, including many examples. Both of them use the \pkg{testthat}  system for unit testing \citep{testthat}, to make the software development more safe and stable and currently reach a code coverage of roughly $95\%$.

As potential future extensions, the \pkg{funData} package could also include \code{irregFunData} objects with observation points in a higher dimensional space or provide appropriate plotting methods for one-dimensional curves in 2D or 3D space. For the \pkg{MFPCA}, new basis functions as e.g.,\ wavelets could be implemented.

\section*{Acknowledgements}

The author thanks two anonymous reviewers for their insightful and constructive comments that definitively helped to improve the paper and the described packages.

\bibliographystyle{apalike}
{
\small
\bibliography{funSoftware}
}

\newpage

\section*{Appendix}

\section*{\pkg{MFPCA}: Additional Plots for the Case Study}

\begin{minipage}{\textwidth}
\begin{center}
\includegraphics[width = 0.45\textwidth]{\PathSoft/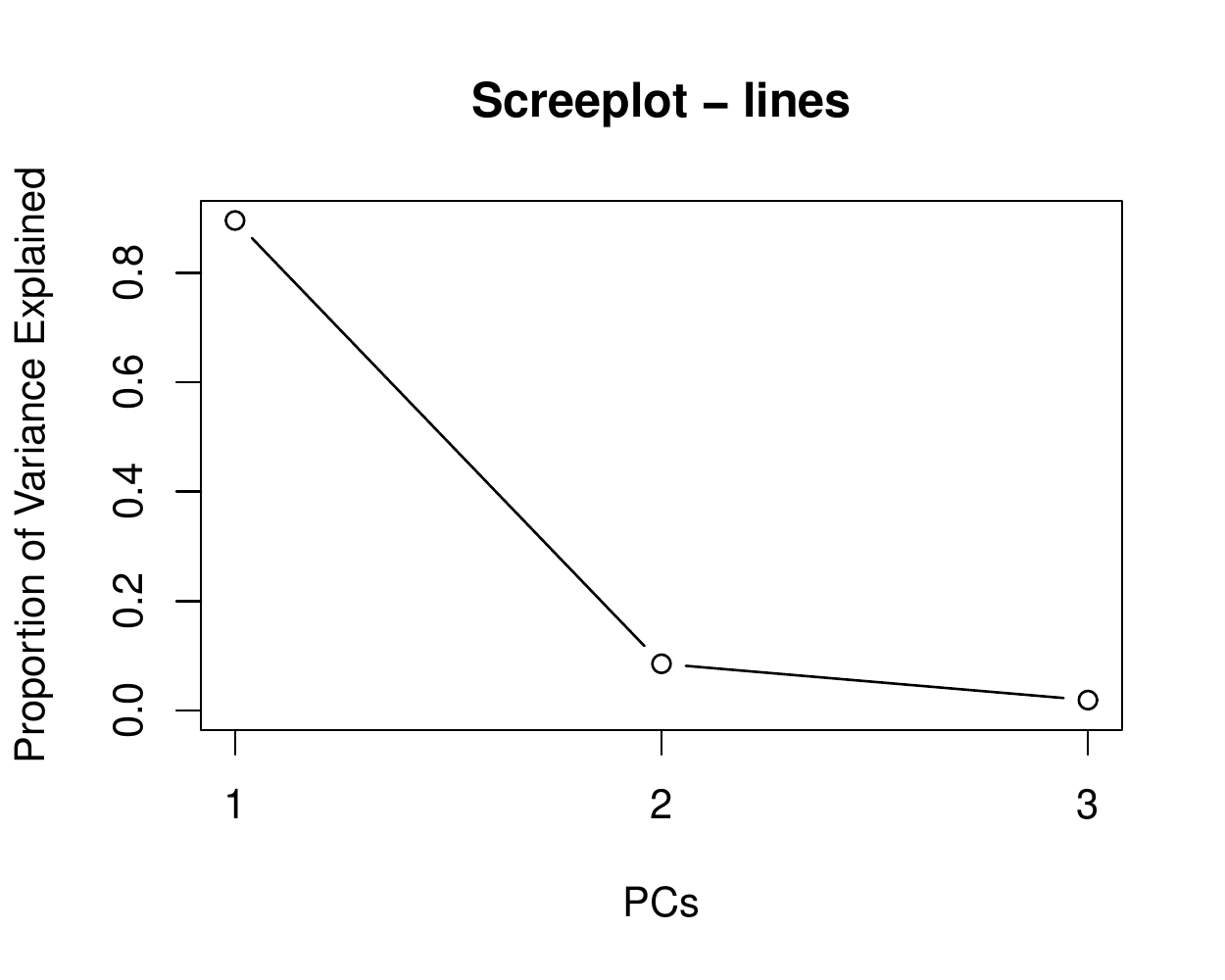}
\includegraphics[width = 0.45\textwidth]{\PathSoft/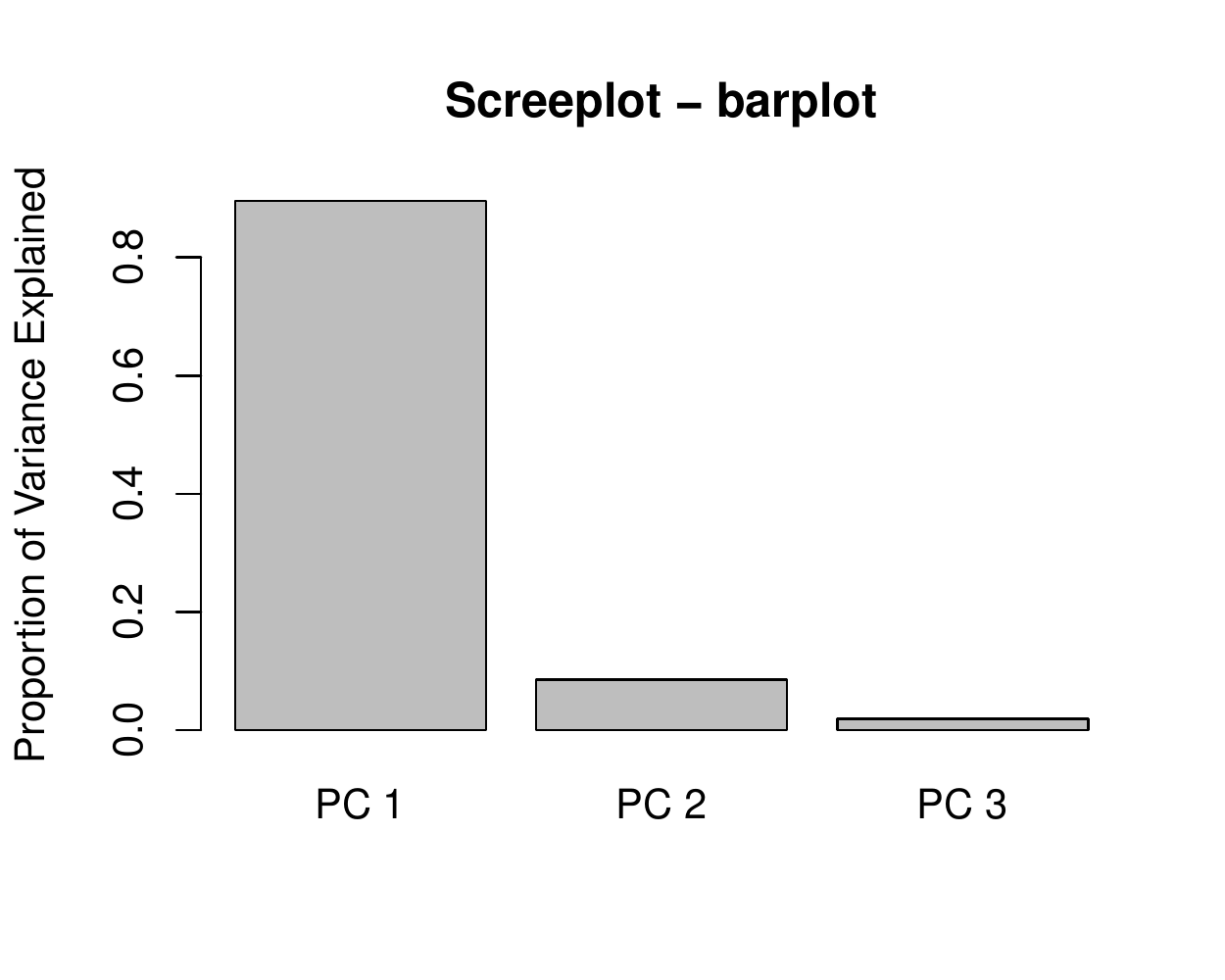}
\end{center}
\captionof{figure}{Screeplots for visualizing the decrease of the eigenvalues for an \code{MFPCAfit} object, here obtained via \code{screeplot(MFPCAweather)}. Left: The default plot (option \code{type = "lines"}). Right: The barplot version (\code{type = "barplot"}).}
\label{fig:MFPCAscreeplot}
\end{minipage}

\begin{minipage}{\textwidth}
\begin{center}
\includegraphics[width = 0.85\textwidth]{\PathSoft/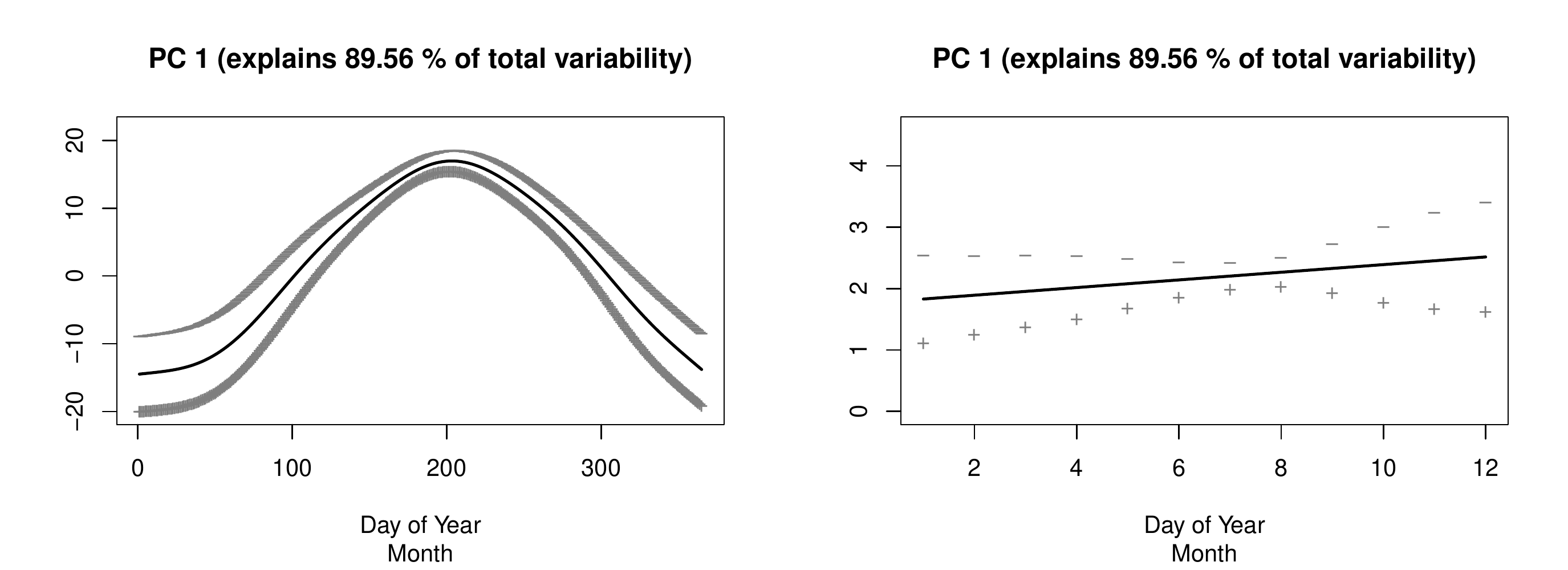}
\end{center}
\captionof{figure}{The first principal component of the Canadian weather data as perturbation of the mean via \code{plot(MFPCAweather, combined = TRUE)}. The plots show the effects of adding ('+') and subtracting ('-') a multiple of the principal component to the bivariate mean function.}
\label{fig:plotMFPCAfit}
\end{minipage}

\begin{minipage}{\textwidth}
\begin{center}
\includegraphics[width = 0.6\textwidth]{\PathSoft/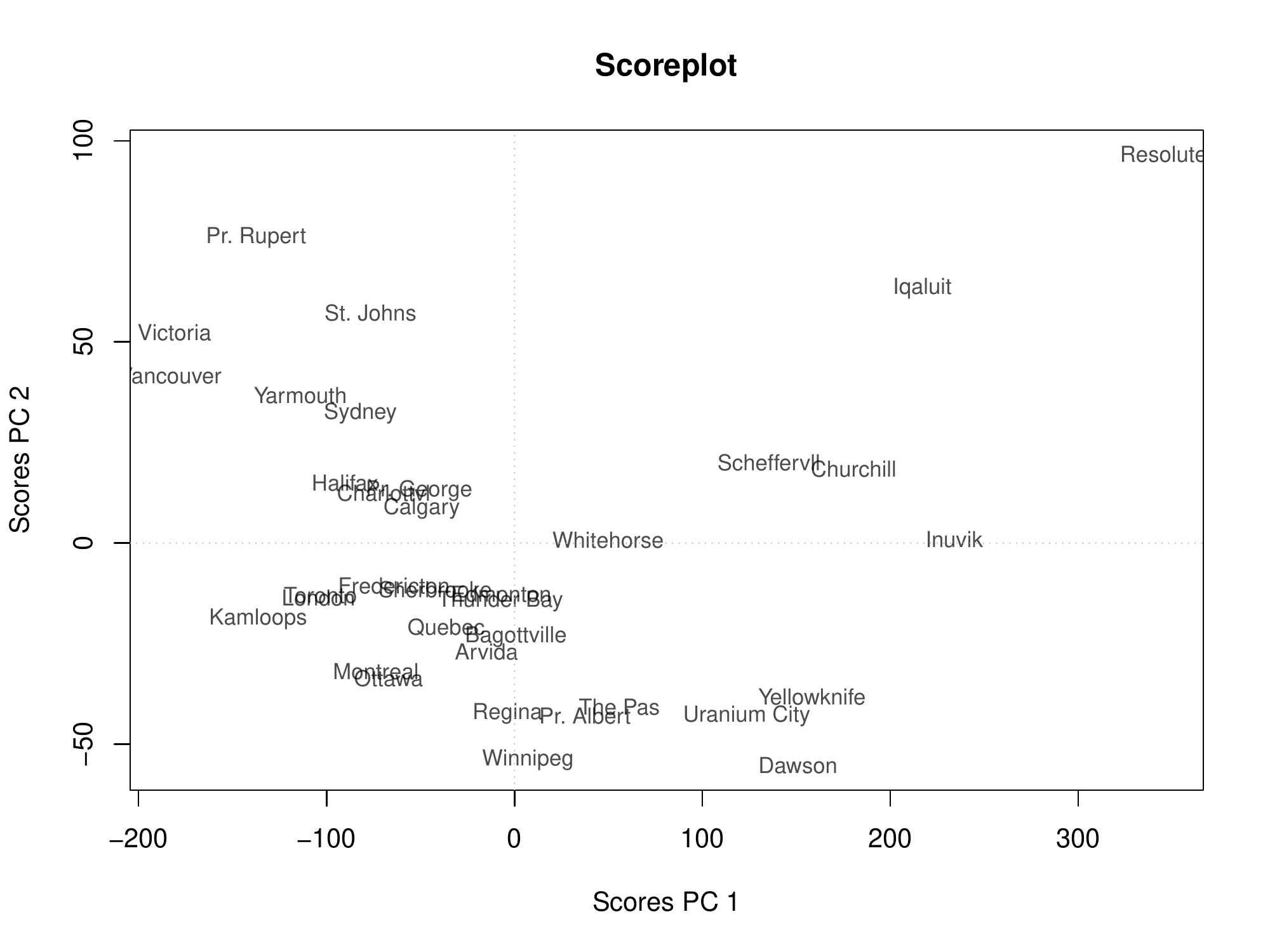}
\end{center}
\captionof{figure}{The scores of the first two principal components of the Canadian weather data plotted via \code{scoreplot(MFPCAweather)}. The labels give the names of the $35$ weather stations.}
\label{fig:plotMFPCAscores}
\end{minipage}

\section*{\pkg{MFPCA}: Univariate basis expansions}

\code{given}: Given basis functions. This can for example be useful if a univariate FPCA was already calculated for each element. For one element, \code{uniExpansions} looks as follows:
\begin{Schunk}
\begin{Sinput}
R> list(type = "given", functions, scores, ortho)
\end{Sinput}
\end{Schunk}
Here \code{functions} is a \code{funData} object on the same domain as the data and contains the given basis functions. The parameters \code{scores} and \code{ortho} are optional. The first represents the coefficient matrix of the observed functions for the given basis functions in a row-wise manner, while \code{ortho} specifies whether the basis functions are orthonormal or not.  If \code{ortho} is not supplied, the functions are treated as non-orthonormal.

\bigskip

\code{uFPCA}: Univariate functional principal component analysis for data on one-dimensional domains. This option was used in the previous example. The list entry for one element has the form:
\begin{Schunk}
\begin{Sinput}
R> list(type = "uFPCA", nbasis, pve, npc, makePD, cov.weight.type)
\end{Sinput}
\end{Schunk}
The implementation is based on the PACE approach \citep{YaoEtAl:2005} with the mean function and the covariance surface smoothed with penalized splines \citep{DiEtAl:2009}, following the implementation in the \pkg{refund} package. The \code{MFPCA} function returns the smoothed mean function, while for all other options, the mean function is calculated pointwise. Options for this expansion include the number of basis functions \code{nbasis} used for the smoothed mean and covariance functions (defaults to 10; for the covariance this number of basis functions is used for each marginal); \code{pve}, a value between $0$ and $1$, giving  the proportion of variance that should be explained by the principal components (defaults to $0.99$); \code{npc}, an alternative way to specify the number of principal components to be calculated explicitly (defaults to \code{NULL}, otherwise overrides \code{pve}); \code{makePD}, an option to enforce positive definiteness of the covariance surface estimate (defaults to \code{FALSE}) and \code{cov.weight.type}, which characterizes the weighting scheme for the covariance surface (defaults to \code{"none"}).

\bigskip

\code{spline1D} and \code{spline1Dpen}: These options calculate a spline representation of functions on one-dimensional domains using the \code{gam} function in the \pkg{mgcv} package \citep{Wood:2011, mgcv}. When using this option, the \code{uniExpansions} entry for one element is of the form:
\begin{Schunk}
\begin{Sinput}
R> list(type = "splines1D", bs, m, k)
R> list(type = "splines1Dpen", bs, m, k, parallel)
\end{Sinput}
\end{Schunk}
 For \code{spline1Dpen}, the coefficients are found by a penalization approach, while for \code{spline1D} the observations are simply projected on the spline space without penalization. Thus, the \code{spline1Dpen} option will in general lead to smoother representations than \code{spline1D}. Possible options passed for these expansions are
\code{bs}, the type of basis functions to use (defaults to \code{"ps"} for possibly penalized B-spline functions); \code{m}, the order of the spline basis (defaults to \code{NA}, i.e., the order is chosen automatically); \code{k}, the number of basis functions to use (default value is \code{-1}, which means that the number of basis functions is chosen automatically). For the penalized version, there is an additional option \code{parallel} which, if set to \code{TRUE}, calculates the spline coefficients in parallel. In this case, a parallel backend must be registered before (defaults to \code{FALSE}).

\bigskip

\code{spline2D} and \code{spline2Dpen}: These are analogue options to \code{spline1D} and \code{spline1Dpen} for functional data on two-dimensional domains (images):
\begin{Schunk}
\begin{Sinput}
R> list(type = "splines2D", bs, m, k)
R> list(type = "splines2Dpen", bs, m, k, parallel)
\end{Sinput}
\end{Schunk}
 The parameters  \code{bs}, \code{m} and \code{k} for the type, order and number of basis functions can be either a single number/character string that is used for all marginals or a vector with the specifications for all marginals. For the penalized version, the function \code{bam} in \pkg{mgcv} is used to speed up the calculations and reduce memory load. Setting \code{parallel=TRUE} enables parallel calculation of the basis function coefficients. As for the one-dimensional case, this requires a parallel backend to be registered before.

\bigskip

\code{fda}: This option allows to use all basis functions expansions implemented in the package \pkg{fda}, such as for example the leading $15$ basis functions of the Fourier basis on $[0,1]$:
\begin{Schunk}
\begin{Sinput}
R> basis <- fda::create.fourier.basis(c(0,1), nbasis=15)
R> list(type = "fda", basis)
\end{Sinput}
\end{Schunk}
All parameters are passed to the coercion method \code{funData2fd}, which heavily builds on the function \code{eval.fd} from the \pkg{fda} package. If this package is not available, an error is thrown and the calculation is stopped.
 
 \bigskip

\code{FCP\_TPA}: This option uses the \textit{Functional CP-TPA} algorithm of \citet{Allen:2013} to compute an eigendecomposition of image observations, which can be interpreted as functions on a two-dimensional domain. The algorithm assumes a CANDECOMP/PARAFRAC (CP) representation of the data tensor $X \in \IR^{N \times S_x \times S_y}$ containing all observations $x_i$ with $S_x \times S_y$ pixels, each:
\[X = \sum_{m = 1}^M d_m u_m \circ v_m \circ w_m\]
Here, $d_m$ is a scalar, $u_m \in \IR^{N}, v_m \in \IR^{S_x}, w_m \in \IR^{S_y}$ are vectors and $\circ$ denotes the outer product. We can thus interpret $v_m \circ w_m$ as the $m$-th univariate eigenfunction evaluated at the same pixels as the originally observed data. The vector $d_m \cdot u_m \in \IR^{N}$ can in turn be interpreted as the score vector containing the scores for the $m$-th principal component function and each observation. The algorithm proposed in \citet{Allen:2013} includes smoothing parameters $\lambda_u, \lambda_v, \lambda_w \geq 0$ to smooth along all dimensions, extending the approach of \citet{HuangEtAl:2009} from one-dimensional to two-dimensional functions. As smoothing along the observations $u_m \in \IR^N$ is not required in the given context, the parameter $\lambda_u$ is fixed to zero and the smoothing is implemented only for the $v$ and $w$ directions.
When decomposing images with this algorithm, the user has to supply a list of the following form for the corresponding element:
\begin{Schunk}
\begin{Sinput}
R> list(type = "FCP_TPA", npc, smoothingDegree, alphaRange, 
+       orderValues, normalize)
\end{Sinput}
\end{Schunk}
Required options are \code{npc}, the number of eigenimages to be calculated, and \code{alphaRange}, the range of the smoothing parameters. The latter must be a list with two entries named \code{v} and \code{w}, giving the possible range for $\lambda_v, \lambda_w$ as vectors with the minimal and maximal value, each (e.g.,\ 
\code{alphaRange = list(v = c(10\string^-2,10\string^2), w = c(10\string^-3,10\string^3))}
 would enforce $\lambda_v \in [10^{-2},10^2]$ and $\lambda_w \in [10^{-3}, 10^{3}]$). Optimal  values for $\lambda_v$ and $\lambda_w$ are found by numerically optimizing a GCV criterion \citep[cf.][in the one-dimensional case]{HuangEtAl:2009}. Further options are the smoothing degree, i.e., the type of differences that should be penalized in the smoothing step (\code{smoothingDegree}, defaults to second differences for both directions) and two logical parameters concerning the ordering of the principal components and their normalizations: If  \code{orderValues} is \code{TRUE}, the eigenvalues and associated eigenimages and scores are ordered decreasingly (defaults to \code{TRUE}), i.e., the first eigenimage corresponds to the highest eigenvalue that has been found, the second eigenimage to the second highest eigenvalue and so on. The option \code{normalize} specifies whether the eigenimages should be normalized (defaults to \code{FALSE}).

\bigskip

\code{UMPCA}: This option implements the UMPCA \citep[Uncorrelated Multilinear Principal Component Analysis,][]{LuEtAl:2009} algorithm for finding uncorrelated eigenimages of two-dimensional functions (images). Essentially, this implements the UMPCA toolbox for MATLAB \citep{UMPCA} in \proglang{R}:
\begin{Schunk}
\begin{Sinput}
R> list(type = "UMPCA", npc)
\end{Sinput}
\end{Schunk}
 The number of eigenimages that are calculated has to be supplied by the user (\code{npc}). Note that this algorithm aims more at uncorrelated features than at an optimal reconstruction of the images and thus may lead to unsatisfactory results for the MFPCA approach.

\bigskip

\code{DCT2D}/\code{DCT3D}: This option calculates a representation of functional data on two- or three-di\-men\-sio\-nal domains in a tensor cosine basis. For speeding up the calculations, the implementation is based on the \pkg{fftw3} \proglang{C}-library \citep[][developer version]{fftw3}. If the \pkg{fftw3-dev} library is not available during the installation of the \pkg{MFPCA} package, the \code{DCT2D} and \code{DCT3D} options are disabled and throw an error. After installing \pkg{fftw3-dev} on the system, \pkg{MFPCA} has to be re-installed to activate \code{DCT2D}/\code{DCT3D}. The \code{uniExpansions} entry for a cosine representation of 2D/3D elements is:
\begin{Schunk}
\begin{Sinput}
R> list(type = "DCT2D", qThresh, parallel)
R> list(type = "DCT3D", qThresh, parallel)
\end{Sinput}
\end{Schunk} 
The discrete cosine transformation is a real-valued variant of the fast Fourier transform (FFT) and usually results in a huge number of non-zero coefficients that mostly model ``noise'' and can thus be set to zero without affecting the representation of the data. The user has to supply a threshold between $0$ and $1$ (\code{qThresh}) that defines the proportion of coefficients to be thresholded. Setting e.g.,\ \code{qThresh = 0.9} will set $90\%$ of the coefficients to zero, leaving only the $10\%$ of the coefficients with the highest absolute values. The coefficients are stored in a \code{sparseMatrix} (package \pkg{Matrix}) object to reduce the memory load for the following computations. The calculations can be run in parallel for the different observations by setting the parameter \code{parallel} to \code{TRUE} (defaults to \code{FALSE}), if a parallel backend has been registered before.

\end{document}